\documentclass[a4paper,12pt]{article}
\pdfoutput=1

\usepackage{jcappub} 

\usepackage{subfigure}
\usepackage{dcolumn}
\usepackage{array} 
\usepackage{longtable}
\usepackage{booktabs} 
\usepackage[T1]{fontenc} 
\usepackage{hyperref}
\usepackage{multirow}
\usepackage{caption}

\newcommand{\nn}{\nonumber}
\def\al{\alpha}

\def\Om{\Omega}

\def\doi{http://doi.org}

\def\m{\mathrm{m}}

\def\d{\mathrm{d}}

\usepackage{xcolor}

\usepackage{orcidlink}

\title{Beyond CPL: Evidence for Dynamical Dark Energy in Three-Parameter Models}

\author[a]{Sonej Alam\orcidlink{0009-0008-8322-2923}}
\author[a]{Md.~Wali Hossain\orcidlink{0000-0001-6969-8716}}

\affiliation[a]{Department of Physics, Jamia Millia Islamia,\\ New Delhi 110025, India}

\emailAdd{sonejalam36@gmail.com}
\emailAdd{mhossain@jmi.ac.in}

\abstract{We introduce two three-parameter extensions of the minimal Akhtar–Hossain (mAH) dark energy parametrization, termed modified minimal AH (MmAH1 and MmAH2), which provide a smooth and bounded evolution of the dark energy equation of state while retaining $\Lambda$CDM as a limiting case. Using a joint analysis of the CMB compressed likelihood, DESI DR2 BAO, $H(z)$, redshift space distortions, and three SNeIa samples (PantheonPlus, Union3, and DESY5), we compare these models with $\Lambda$CDM, $w$CDM, mAH, CPL, and the three-parameter CPL-$w_{\rm b}$ extension. The standard cosmological parameters remain stable across all models, while CPL, MmAH1 and MmAH2 parametrizations yield modest but consistent improvements in fit ($\Delta\chi^2\simeq-6$ to $-12$ for PantheonPlus and Union3, and $\simeq-38$ for DESY5). Statistical consistency with $\Lambda$CDM, quantified via the Mahalanobis distance in one, two, and three dimensional parameter subspaces, reveals mild to moderate deviations, $\sim2$--$2.5\sigma$ for $+$PantheonPlus, $2$--$3\sigma$ for $+$Union3, and up to $4$--$5\sigma$ for $+$DESY5 combination, depending on model complexity. Among all extensions CPL, MmAH1 and MmAH2 provide the most stable and physically coherent representations of dynamical dark energy, maintaining moderate tensions with $\Lambda$CDM and well behaved parameter correlations. Overall, these results indicate consistent evidence for departures from $\Lambda$CDM.}


\begin{document}
\maketitle
\flushbottom

\section{Introduction}
\label{sec:intro}
Over the past few years, multiple independent datasets have revealed significant tensions that challenge the internal consistency of $\Lambda$ cold dark matter ($\Lambda$CDM), $\Lambda$ being the cosmological constant (CC). The most prominent is the $>5\sigma$ discrepancy between the local Hubble parameter at present ($H_0$) measurement by Supernovae (SNe) and $H_0$ for the Equation of State of dark energy (SH0ES) team~\cite{Riess:2021jrx} and the Cosmic microwave background (CMB)-inferred value under $\Lambda$CDM~\cite{Planck:2018vyg}, commonly known as the Hubble tension. Another is the $S_8=\sigma_8\sqrt{\Omega_{\rm m0}/0.3}$ tension~\cite{Perivolaropoulos:2021jda,Kilo-DegreeSurvey:2023gfr}, where $\sigma_8$ is the present-day amplitude of linear matter fluctuations at $8h^{-1}\,{\rm Mpc}$. These discrepancies may originate from unknown systematics or new physics beyond $\Lambda$CDM~\cite{Kamionkowski:2022pkx,Freedman:2023jcz,Bernal:2016gxb,Knox:2019rjx}. More recently, baryon acoustic oscillations (BAO) measurements by the Dark Energy Spectroscopic Instrument (DESI) has reported hints of dynamical dark energy (DDE), first data release (DR1) found a $2.6\sigma$ \cite{DESI:2024mwx} preference for DDE over $\Lambda$CDM in DESI$+$CMB using the Chevallier-Polarski-Linder (CPL) parametrization \cite{Chevallier:2000qy,Linder:2002et}, which increased to $3.1\sigma$ in DR2~\cite{DESI:2025zgx,DESI:2025wyn}. The inclusion of various Type Ia supernova (SNeIa) samples further strengthens this preference, to the $2.8\sigma$--$4.2\sigma$ level~\cite{DESI:2025zgx,DESI:2024aqx,DESI:2024kob,DESI:2025wyn}. These findings have stimulated a surge of theoretical and phenomenological studies on dynamical dark energy~\cite{Giare:2024smz,Berghaus:2024kra,Wolf:2025jlc,Lee:2025pzo,Zhong:2025gyn,Yao:2025wlx,Qu:2024lpx,Wang:2024dka,Giare:2024gpk,Gialamas:2024lyw,Shlivko:2024llw,Ye:2024ywg,Bhattacharya:2024hep,Ramadan:2024kmn,Jiang:2024xnu,Payeur:2024dnq,Malekjani:2024bgi,Wolf:2023uno,Wolf:2024eph,Wolf:2024stt,Chan-GyungPark:2024mlx,Park:2024vrw,Park:2024pew,Dinda:2024kjf,Dinda:2024ktd,Jiang:2024viw,Colgain:2024xqj,Bhattacharya:2024kxp,Akthar:2024tua,Chan-GyungPark:2025cri,Ferrari:2025egk,Wolf:2025jlc,Peng:2025nez,Colgain:2024mtg,Mukherjee:2024ryz,Mukherjee:2025myk,Mukherjee:2025ytj,RoyChoudhury:2024wri,Berbig:2024aee,Li:2024qso,Li:2024qus,Li:2025owk,Borghetto:2025jrk,Carloni:2024zpl,Liu:2025mub,Wolf:2025jed,Chudaykin:2024gol,Sohail:2024oki,Hossain:2025grx,Cheng:2025lod,Scherer:2025esj,Pedrotti:2025ccw,Escamilla:2023oce,Liu:2025myr,Wolf:2025acj,Cline:2025sbt,Wang:2025dtk,VanRaamsdonk:2025wvj,Peng:2025tqt,Lu:2025gki,Yang:2025oax,Jiang:2025hco,Ishak:2025cay,Shlivko:2025fgv,Silva:2025twg,Gialamas:2025pwv,Cortes:2025joz,Artola:2025zzb,Dinda:2025hiu,Park:2025fbl,Blanco:2025vva,Chaudhary:2025pcc,Chaudhary:2025vzy,Li:2025muv,Zhang:2025dwu}.

A widely used framework for probing dynamical dark energy (DDE) is the two parameter ($w_0,w_{\rm a}$) CPL parametrization~\cite{Chevallier:2000qy,Linder:2002et} of the dark energy equation of state (EoS), $w(a) = w_0 + w_{\rm a}(1-a)$, which corresponds to a first-order Taylor expansion of $w(a)$ around $a=1$, where $a$ is the scale factor. While the CPL form is economical and analytically tractable, it may be too restrictive to capture possible transitions at higher redshifts. One way to improve flexibility is to add quadratic terms in CPL such as $w_{\rm b}(1-a)^2$ (CPL-$w_{\rm b}$), or higher-order terms of the expansion, though this comes at the cost of additional free parameters that may be poorly constrained by current data \cite{Nesseris:2025lke}.

To explore dark energy dynamics beyond CPL, the Akthar-Hossain (AH) parametrization~\cite{Akthar:2024tua} has been proposed as a general framework for minimally coupled canonical scalar fields, capable of reproducing thawing, scaling-freezing, and tracker behaviours at any redshift, and even allowing a phantom regime at late times. For thawing and phantom dynamics, AH admits a minimal version (mAH) where the scalar field EoS approaches $-1$ at large redshifts, a restrictive feature, and statistical analyses show that mAH is not preferred over CPL or $w$CDM~\cite{Akthar:2024tua}. To address this, we propose an extension by promoting the high-redshift asymptote of the EoS to a free parameter, defining the modified minimal AH (MmAH) parametrization. While mAH has two free parameters like CPL, MmAH introduces a third, offering greater flexibility in describing dark energy dynamics.

The primary objective of this work is to investigate observational evidence for DDE, with particular emphasis on the proposed MmAH1 and MmAH2 parametrizations. To this end, we constrain the cosmological parameters of seven representative models, (i) the standard $\Lambda$CDM model, (ii) the $w$CDM model with a constant dark energy EoS $w$, (iii) the CPL parametrization characterized by $(w_0, w_{\rm a})$, (iv) its three-parameter extension CPL-$w_{\rm b}$ with $(w_0, w_{\rm a}, w_{\rm b})$, (v) the mAH form, and its two modified variants, (vi) MmAH1 and (vii) MmAH2. Parameter estimation is carried out using Markov Chain Monte Carlo (MCMC) techniques, combining multiple cosmological probes, the CMB distance priors from \textit{Planck} 2018 TT, TE, EE$+$lowE~\cite{Planck:2018vyg,2019JCAP}, DESI DR2 BAO measurements~\cite{DESI:2025zgx}, Hubble parameter data~\cite{2018JCAP...04..051G}, redshift-space distortion (RSD) measurements~\cite{Nesseris:2017vor} and three Type Ia supernova (SNeIa) samples — PantheonPlus~\cite{Scolnic:2021amr,Brout:2022vxf}, Union3~\cite{Rubin:2023ovl}, and the Dark Energy Survey Year~5 (DESY5) compilation~\cite{DES:2024jxu}. We compare these models using the Akaike Information Criterion (AIC)~\cite{Akaike74} and the Bayesian Information Criterion (BIC)~\cite{Schwarz:1978tpv}. In addition, we quantify their statistical consistency with $\Lambda$CDM by computing parameter space tensions through the Mahalanobis distance~\cite{Mahalanobis:1936} in one, two, and three dimensional subspaces, thereby assessing the overall evidence for DDE.

We review the mAH parametrization in Sec.~\ref{sec:mAH} and introduce its extensions (MmAH1,2) parametrizations, in Sec.~\ref{sec:model}. The methodology of the data analysis is described in Sec.~\ref{sec:data_analysis}, and the cosmological data sets employed are summarized in Sec.~\ref{sec:data}. Results are presented in Sec.~\ref{sec:results}, followed by a discussion of the statistical tension with the standard $\Lambda$CDM model in Sec.~\ref{sec:tension}. Finally, we conclude in Sec.~\ref{sec:conc}.

\section{Review of \lowercase{m}AH Parametrization}
\label{sec:mAH}

The mAH EoS, designed to represent thawing-like and phantom evolution, is given by
\begin{equation}
    w_{\rm mAH}(z) \;=\; -1 + \frac{\alpha}{1 + \left( \dfrac{1+z}{\Omega_\delta} \right)^{3\alpha}},
    \label{eq:AH}
\end{equation}
where $\alpha$ controls the amplitude and direction of deviation from $\Lambda$CDM, and $\Omega_\delta$ determines the transition scale between the matter-dominated and dark-energy-dominated epochs.

$w_{\rm mAH}$ smoothly evolves from $-1$ at high redshift towards a less negative value $-1 + \alpha$ in the future. A transition redshift, $z_t$, defined by $1+z_t = \Omega_\delta$, marks the epoch where $ w_{\rm mAH}(z_t) \;=\; -1 + \frac{\alpha}{2}$. For $\alpha>0$, the model exhibits a thawing behaviour with $w_{\rm mAH}>-1$ while for $\alpha<0$, it allows a phantom dynamics with $w_{\rm mAH}<-1$ at late times. 

The effective CPL parameters corresponding for mAH are
\begin{align}
    w_0^{\rm mAH} &= w(0)= -1 + \frac{\alpha}{1 + \Omega_\delta^{-3\alpha}} \\
    w_a^{\rm mAH} &= -\frac{\d w(a)}{\d a}\Bigg|_{a=1} = - \frac{3 \alpha^2 \, \Omega_\delta^{3\alpha}}{(1 + \Omega_\delta^{3\alpha})^2}
\end{align}
with $w_a^{\rm mAH} < 0$ for all real $\alpha$. Thus, in the mAH parametrization, the slope of the EoS near $z=0$ is always negative, irrespective of whether the model is phantom-like or non-phantom.

mAH has some limitations: (i) it forbids phantom crossing, (ii) in the phantom regime, the EoS becomes increasingly negative in the future, making it unsuitable for scenarios where $w$ stabilizes or evolves back toward $-1$; (iii) at high redshifts the model always approaches $w\simeq -1$, restricting its ability to describe diverse early-time behaviours; and (iv) since $w_a^{\rm mAH}<0$ always, it cannot represent models where the EoS becomes more negative with increasing redshift. Thus, we look for a more general parametrization.

\section{Modified minimal AH}
\label{sec:model}

To allow greater flexibility, we consider the following two modifications of the mAH parametrization~\eqref{eq:AH},
\begin{eqnarray}
     w(z) &=& \beta + \frac{\alpha}{1 + \dfrac{1+z}{\Omega_\delta}} \, ,
    \label{eq:MmAH} \\
     w(z) &=& \beta + \frac{\alpha z}{1 + \dfrac{1+z}{\Omega_\delta}} \, .
    \label{eq:wz-main}
\end{eqnarray}
Here, $\beta$ is an additional free parameter. These two forms can be regarded as special subclasses of a more general parametrization,
\begin{equation}
    w(z) \;=\; \beta \;+\; \frac{\alpha \, z^\xi}{1 + \left(\dfrac{1+z}{\Omega_\delta}\right)^\zeta},
    \label{eq:Para_gen}
\end{equation}
where $\xi$ and $\zeta$ are constants. For $\beta=-1$, $\xi=0$ and $\zeta=3\alpha$, the above parametrization reduces to the mAH case~\eqref{eq:AH}.  
In this work, we restrict our attention to the two simple modifications in Eqs.~\eqref{eq:MmAH} and \eqref{eq:wz-main}, which we refer to as MmAH1 and MmAH2, respectively.

\subsection{MmAH1}
For the parametrization~\eqref{eq:MmAH} (MmAH1), the limiting values are
\begin{align}
    w(z\to \infty) &= \beta, \\
    w(z\to -1) &= \beta+\alpha.
\end{align}
At the transition redshift $1+z_t=\Omega_\delta$, we obtain
\begin{equation}
    w(z_t) = \beta + \frac{\alpha}{2}.
\end{equation}
Expansion around $z=0$ yields CPL-like parameters
\begin{align}
    w_0 = \beta + \frac{\alpha\Omega_\delta}{1+\Omega_\delta},  \;  
    w_a &= -\frac{\alpha\Omega_\delta}{(1+\Omega_\delta)^2}, \; 
    w_b = -\frac{\alpha\Omega_\delta^2}{(1+\Omega_\delta)^3}.
\end{align}
The corresponding dark energy density evolution is
\begin{equation}
    \rho_{\rm DE}(z) = \rho_{\rm DE,0}
    (1+z)^{3(1+\beta+\alpha)}
    \left(
      \frac{1+\Omega_\delta}{\Omega_\delta+1+z}
    \right)^{3\alpha},
    \label{eq:rhoDE_MmAH1}
\end{equation}
where $\rho_{\rm DE,0}$ denotes the present value of the dark energy density.

Defining the equality redshift $z_{\rm eq}$ by
$\rho_\m(z_{\rm eq})=\rho_{\rm DE}(z_{\rm eq})$ we can represent the transition redshift $z_{\rm t}$ in terms of $z_{\rm eq}$ as
\begin{equation}
\Om_\delta=1+z_{\rm t} =
\frac{
(1+z_{\rm eq})\left\{\left(\dfrac{\Omega_{\m0}}{\Omega_{\rm DE0}}\right)^{\tfrac{1}{3\alpha}}
- (1+z_{\rm eq})^{\tfrac{\beta}{\alpha}}\right\}
}{
(1+z_{\rm eq})^{1+\tfrac{\beta}{\alpha}} -
\left(\dfrac{\Omega_{\m0}}{\Omega_{\rm DE0}}\right)^{\tfrac{1}{3\alpha}}
}.~~
\label{eq:Omega_delta_zeq}
\end{equation}
Because of the above analytic form we consider $z_{\rm eq}$ as a free parameter instead of $\Omega_\delta$.

\subsection{MmAH2}
For the parametrization~\eqref{eq:wz-main} (MmAH2), we note that
\begin{align}
    w(z\to \infty) &= \beta+\alpha\Omega_\delta, \\
    w(z\to -1) &= \beta - \alpha.
\end{align}
At the transition scale $1+z_t=\Omega_\delta$, we find
\begin{equation}
    w(z_t) = \beta + \frac{\alpha z_{\rm t}}{2},
\end{equation}
which lies midway between the two asymptotic values above.  
Expanding Eq.~\eqref{eq:wz-main} around $z=0$, the effective CPL parameters are
\begin{align}
    w_0 &= \beta, &
    w_a &= \frac{\alpha\Omega_\delta}{1+\Omega_\delta}, &
    w_b &= \frac{\alpha \, \Omega_\delta^2}{(1+\Omega_\delta)^2}.
\end{align}
Unlike the original mAH parametrization~\eqref{eq:AH}, where $w_a^{\rm mAH}<0$ always, the modified models~\eqref{eq:MmAH} and \eqref{eq:wz-main} allow both positive and negative values of $w_a$, thereby enabling a wider range of possible dark energy dynamics. 

The dark energy density, in MmAH2 model, evolves as
\begin{equation}
    \rho_{\rm DE}(z) = \rho_{\rm DE,0}
    (1+z)^{3(1+\beta-\alpha)}
    \left(
      \frac{\Omega_\delta+1+z}{\Omega_\delta+1}
    \right)^{3\alpha(1+\Omega_\delta)},
    \label{eq:rhoDE_MmAH2}
\end{equation}

\begin{figure}[ht]
\centering
\includegraphics[scale=0.42]{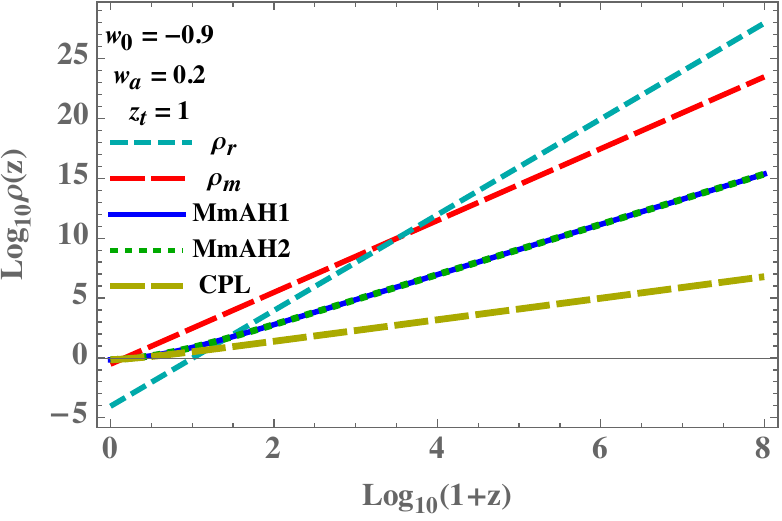}~~~~~~~~~
\includegraphics[scale=0.42]{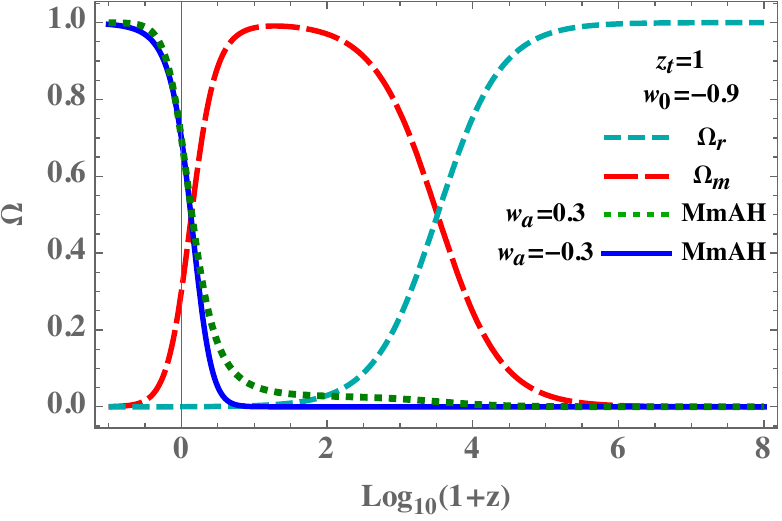} \\
\includegraphics[scale=0.42]{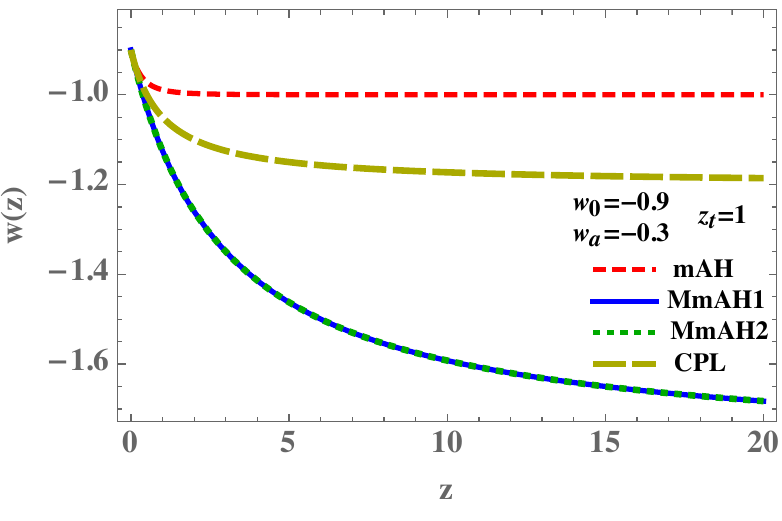}~~~~~~~~~
\includegraphics[scale=0.42]{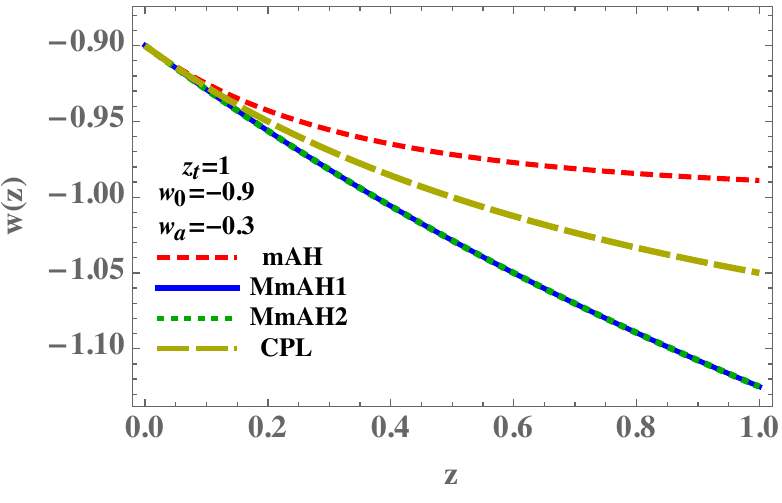}
\caption{Top left: Energy density $\rho$ is plotted against redshift $z$ in $log$ scale for mAH, MmAH1, MmAH2 and CPL.Top right: Evolution of density parameter ($\Omega$) is shown in for MmAH1 and MmAH2 parametrizations along with the evolution of matter and radiation. Bottom left: EoS ($w(z)$) is plotted against $z$. Bottom right: Comparison with CPl in the evolution of EoS is shown for very small redshifts . }
\label{fig:dyn}
\end{figure}

\begin{figure}[ht]
\centering
\includegraphics[height=5cm, width=0.4\linewidth]{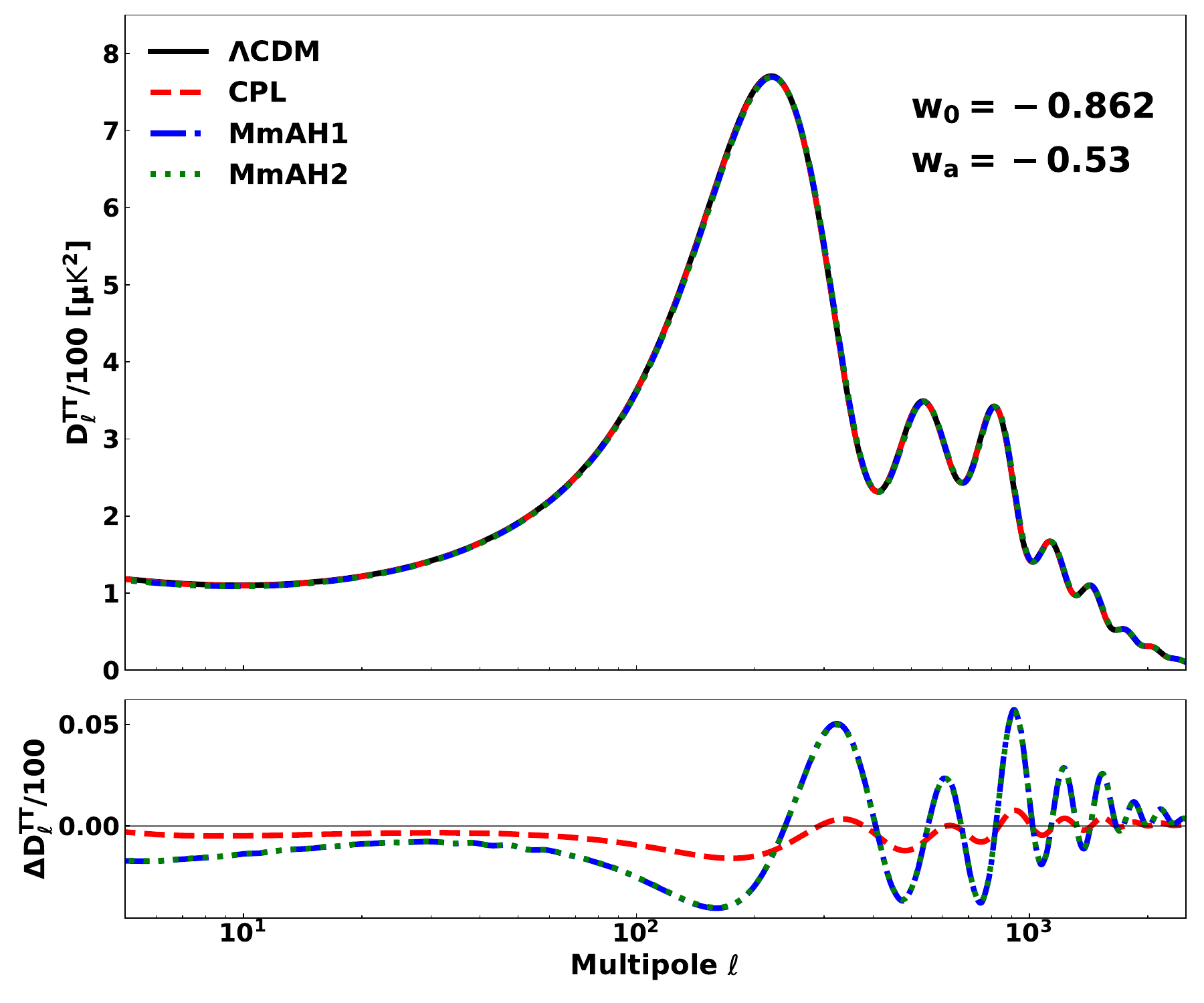}~~~~~~
\includegraphics[height=5cm, width=0.4\linewidth]{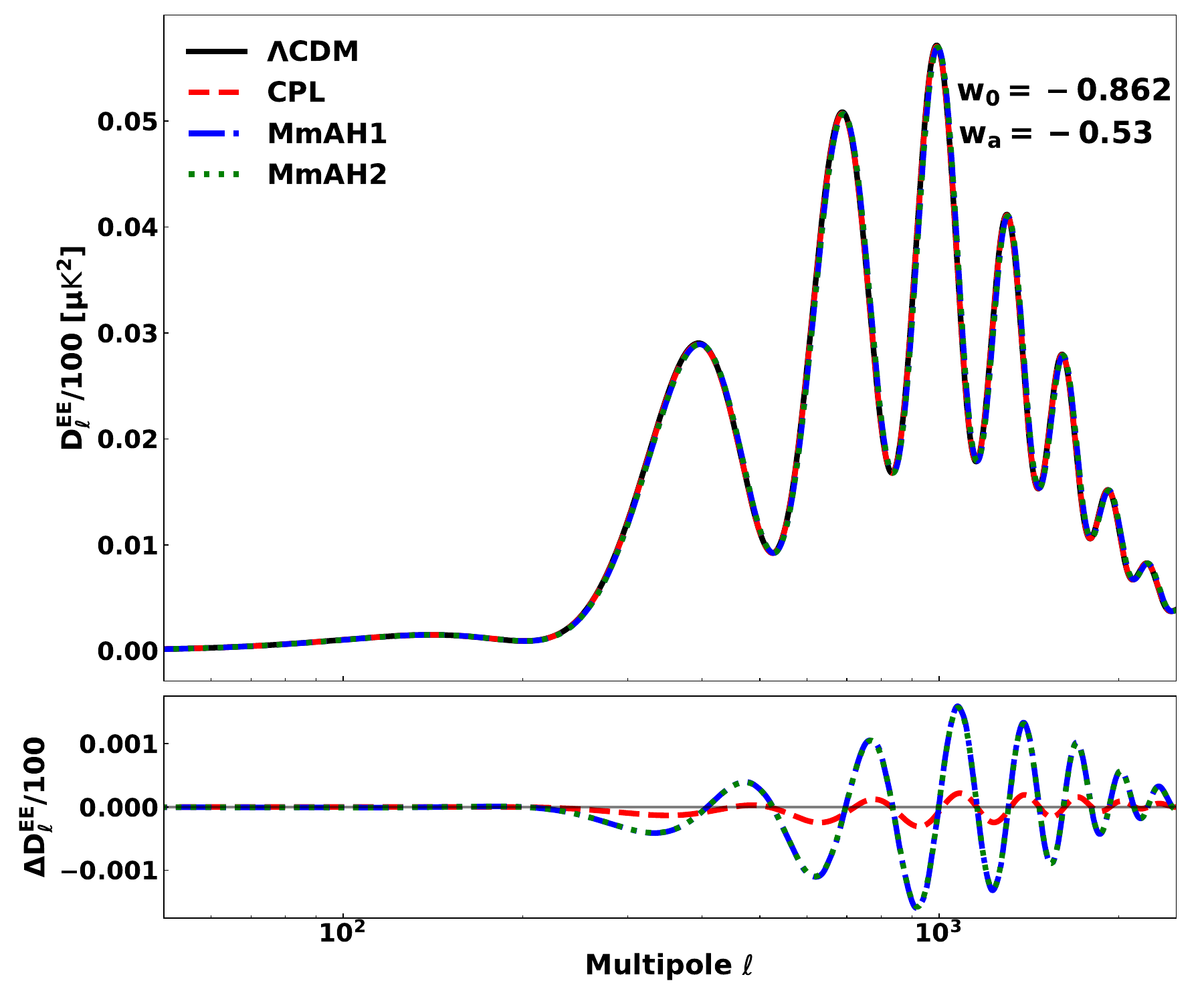}
\includegraphics[height=5cm, width=0.5\linewidth]{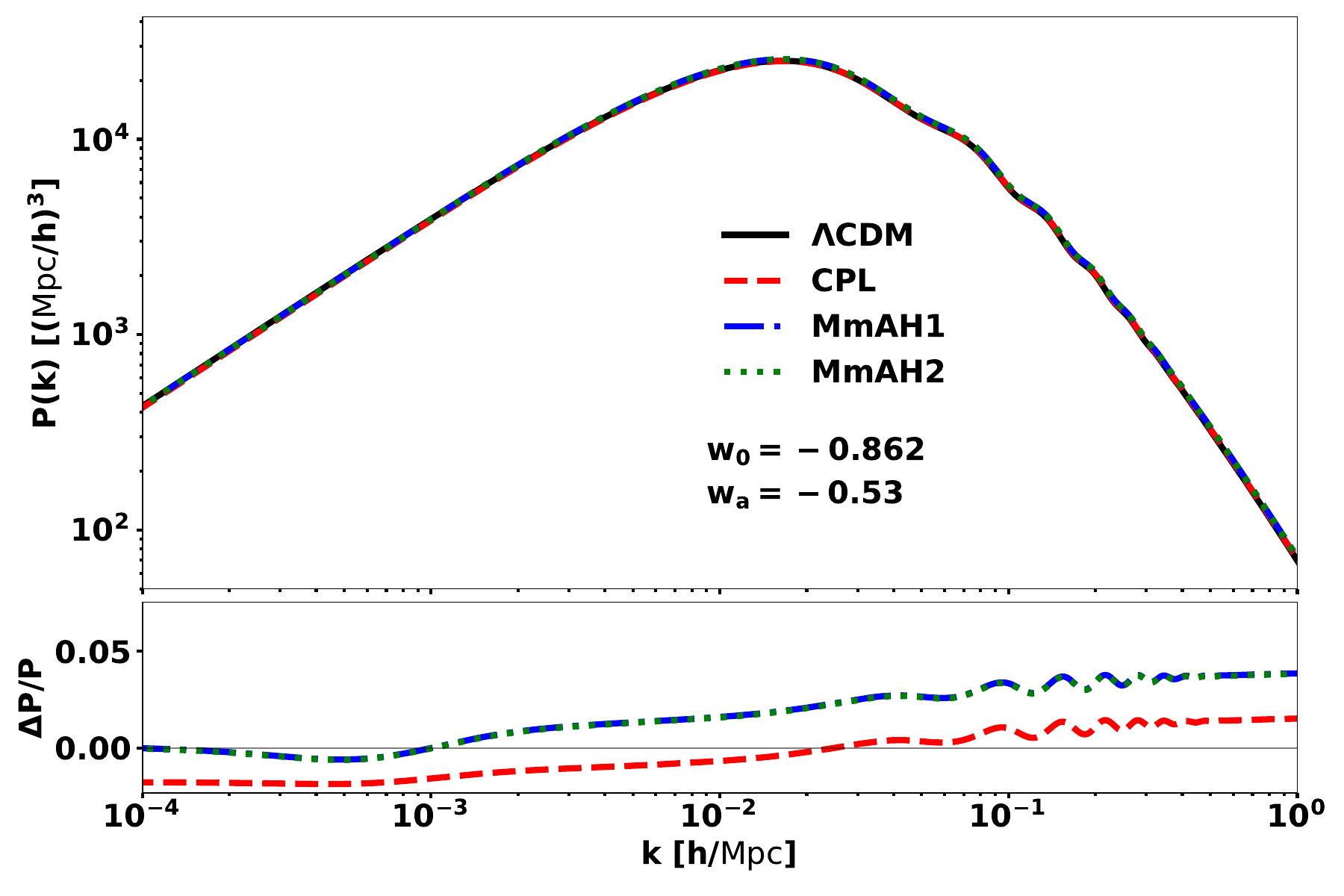}
\caption{Lensed CMB temperature (top) and E-mode polarization (middle) power spectra ($\mathcal{D}_\ell^{TT}$ and $\mathcal{D}_\ell^{EE}$) and their differences with respect to the $\Lambda$CDM, using lensed $C_\ell$ data. 
Bottom: Matter power spectrum $P(k)$ and its difference with respect to the $\Lambda$CDM baseline.}
\label{fig:ps}
\end{figure}

Fig.~\ref{fig:dyn} shows the evolution of the cosmological parameters in different parametrizations. To compare among the models, the figure is plotted using the same values of $w_0$ and $w_{\rm a}$ for all cases. The top-left figure shows the evolution of the energy density of dark energy in different parametrizations, while the top-right figure shows the evolution of the density parameters, $\Omega$, in the MmAH1 and MmAH2 parametrizations. The bottom figures show the evolution of the dark energy EoS in different parametrizations. We find that the MmAH1 and MmAH2 parametrizations exhibit similar cosmological evolutions, which are, however, distinct from those of the mAH and CPL parametrizations. The bottom-right figure shows the behaviour of the EoS, $w(z)$, for redshift up to $z=1$, allowing a comparison with CPL. Since CPL is a Taylor series expansion around $z=0$, we see that mAH, MmAH1, and MmAH2, which are more general parametrizations, overlap with CPL only for very small values of $z<0.1$ for the same $w_0$ and $w_a$ parameters. For $z>0.1$, the evolution of $w(z)$ deviates significantly from its first-order approximation, {\it i.e.}, the CPL parametrization. This shows that keeping only the first-order term in the Taylor expansion makes the evolution of $w(z)$ very restrictive. In this regard, the MmAH1 and MmAH2 parametrizations accommodate a wider range of behaviours in $w(z)$.

Fig.~\ref{fig:ps} shows the CMB TT (top), CMB EE (middle), and matter (bottom) power spectra for different parametrizations with $w_0=-0.862$ and $w_{\rm a}=-0.53$. Additionally, the lower panels of each figure show the deviations with respect to $\Lambda$CDM. We see that, for the same values of $w_0$ and $w_{\rm a}$, all the parametrizations give rise to similar behaviour in the power spectra. The differences in the nature of the power spectra are more clearly visible in the residual plots associated with each case.

\section{Data analysis}
\label{sec:data_analysis}
We examine and constrain seven dark energy models, $\Lambda$CDM, $w$CDM, CPL, CPL-$w_{\rm b}$, mAH, MmAH1 and MmAH2 using the latest cosmological observations. To evaluate and compare the performance of these models we assess their statistical viability using model selection tools \cite{Liddle:2006tc,Shi:2012ma,Trotta:2017wnx}, including the Akaike Information Criterion (AIC) \cite{Akaike74} (${\rm AIC} = 2N + \chi^2_{\rm min}$) and the Bayesian Information Criterion (BIC) \cite{Schwarz:1978tpv} (${\rm BIC} = N \ln k + \chi^2_{\rm min}$), while $k$ and $N$ denoting the total number of data points and free parameters, respectively. Alongside, we report the reduced chi-squared, defined as $\chi_{\rm red}^2 = \chi_{\rm min}^2/\nu$, where $\nu = k - N$ is the number of degrees of freedom. We also compute the relative differences in AIC and BIC with respect to the $\Lambda$CDM, $\Delta {\rm AIC} = \rm AIC_{\rm model} - AIC_{\Lambda \rm CDM}$ and $\Delta {\rm BIC} = \rm BIC_{\rm model} - BIC_{\Lambda \rm CDM}$, which indicate the degree to which a given model is statistically favoured or disfavoured in comparison to the standard model.

\begin{table*}[t]
\begin{center}
\caption{Constraints on the parameters and model comparison statistics for $\Lambda$CDM, $w$CDM, CPL, CPL-$w_b$, mAH, MmAH1, and MmAH2 models using the PantheonPlus, Union3, and DESY5 datasets. All parameter constraints are quoted at the $68\%$ confidence level ($1\sigma$), and upper or lower limits correspond to the same confidence level.}
\label{tab:constraints_combined}
\resizebox{\textwidth}{!}{%
\begin{tabular}{lccccccc}
\hline\hline
\multicolumn{8}{c}{\textbf{+PantheonPlus}}\\
\hline
Parameter & $\Lambda$CDM & $w$CDM & CPL & CPL-$w_b$ & mAH & MmAH1 & MmAH2  \\
\hline
$\Omega_{m0}$ & $0.3087 \pm 0.0059$ & $0.3089 \pm 0.0064$ & $0.3126 \pm 0.0067$ & $0.3097 \pm 0.0067$ & $0.3091 \pm 0.0062$ & $0.3120 \pm 0.0066$ & $0.3116 \pm 0.0066$  \\
$h$ & $0.6789 \pm 0.0045$ & $0.6780 \pm  0.0063$ & $0.6760 \pm 0.0063$ & $0.68 \pm 0.0064$  & $0.6779 \pm 0.0060$ & $0.6770 \pm 0.0062$ & $0.6772 \pm 0.0061$ \\
$\omega_{b}$ & $0.02245 \pm 0.00013$ & $0.02250 \pm 0.00015$ & $0.02241 \pm 0.00015$ & $0.02238 \pm 0.00015$ & $0.02247 \pm 0.00014$ & $0.02241 \pm 0.00015$ & $0.02241 \pm 0.00015$ \\
$r_{\rm d}h$ & $100.39 \pm 0.51$ & $100.20 \pm 0.84$ & $99.69 \pm 0.89$ & $100.04 \pm 0.88$ & $100.22 \pm 0.83$ & $99.69 \pm 0.86$ & $99.78 \pm 0.87$ \\
$\sigma_8$ & $0.75 \pm 0.029$ & $0.750 \pm 0.029$ & $0.752 \pm 0.029$ & $0.751 \pm 0.029$ & $0.751 \pm 0.030$ & $0.752 \pm 0.029$ & $0.752 \pm 0.029$ \\
$w_0$ & \dots & $-0.991 \pm 0.025$ & $ -0.862 \pm 0.059$ & $-1.06^{+0.11}_{-0.13}$ & $-0.993 \pm 0.026$ & $-0.8784 \pm 0.0479$ & $ -0.899 \pm 0.048$\\
$w_a$ & \dots & \dots & $-0.53^{+0.24}_{-0.21}$ & $1.7 ^{+1.3}_{-1.1}$ & $-0.0004\pm 0.028 $ & $-0.3035 \pm 0.1244$ & $ -0.241^{+0.14}_{-0.092}$\\
$w_b$ & \dots & \dots & \dots & $-4 \pm 2.2$ & \dots & $-0.272 \pm 0.1062 $ & $-0.230^{+0.12}_{-0.093}$\\
$\alpha$ & \dots & \dots & \dots  & \dots  & $0.027^{+0.076}_{-0.089}$ & $>0$ & $ -0.257^{+0.15}_{-0.083}$  \\
$z_{\rm t}$ & \dots & \dots  & \dots & \dots & unconstrained & $<40$  & $<134$ \\
$z_{\rm eq}$ & \dots & \dots  & \dots & \dots  & \dots & $0.330 \pm 0.015$ & \dots\\
$\gamma$ & \dots & \dots & \dots & \dots & \dots  &$-0.52^{+0.16}_{-0.26}$ & \dots \\
$\beta$ & \dots & \dots & \dots & \dots & \dots & $<0$ &  $-0.899 \pm 0.048$\\
$M$ & $-19.425 \pm 0.013$ & $-19.426 \pm 0.016$ & $ -19.418 \pm 0.016$ & $-19.417 \pm 0.016$ & $-19.427 \pm 0.015$ & $-19.416 \pm 0.016$ & $-19.417 \pm 0.016$ \\
\hline
$\chi^2_{\rm min}$        & $1447.55$ & $1447.01$ & $1441.57$ & $1444.346$ & $1446.97$ & $1440.39$ & $1440.43$\\
$\chi^2_{\rm red}$        & $0.878$ & $0.878$ & $0.875$ & $0.878$ & $0.879$ & $0.875$ & $0.875$\\
$\Delta\chi^2_{\rm min}$        & $0$ & $-0.54$ & $-5.98$ & $-3.2$ & $-0.58$ & $-7.16$ & $-7.12$ \\
AIC                       & $1459.55$ & $1461.01$ & $1457.57$ & $1462.34$ & $1462.97$ & $1458.39$ & $1458.43$ \\
$\Delta$AIC               & $0$ & $1.46$ & $-1.98$ & $2.79$ & $3.42$ & $-1.16$ & $-1.12$\\
BIC                       & $1492.02$ & $1498.89$ & $1500.86$ & $1511.04$ & $1506.26$ & $1507.08$ & $1507.13$\\
$\Delta$BIC               & $0$ & $6.87$ & $8.84$ & $19.02$ & $14.24$ & $15.06$ & $15.11$\\
\hline
\multicolumn{8}{c}{\textbf{+Union3}}\\
\hline
$\Omega_{m0}$ & $0.3072 \pm 0.0060$ & $0.3073 \pm 0.0070$ & $0.3224 \pm 0.0087$ & $0.3206^{+0.0103}_{-0.0094}$ & $0.3071 \pm 0.0068$ & $0.3210^{+0.0083}_{-0.0087}$ & $0.3184 \pm 0.0085$  \\
$h$ & $0.6800 \pm 0.0047$ & $0.6794 \pm 0.0072$ & $0.666 \pm 0.0082$ & $0.6682^{+0.0093}_{-0.0097}$ & $0.6796 \pm 0.0073$ & $0.6674^{+0.0080}_{-0.0078}$ & $0.6701 \pm 0.0080$ \\
$\omega_{b}$ & $0.02248 \pm 0.00013$ & $0.02250 \pm 0.00014$ & $0.02242 \pm 0.00015$ & $0.022396^{+0.000149}_{-0.000149}$ & $0.02250 \pm 0.00015$ & $0.022408^{+0.000150}_{-0.000152}$ & $0.02243 \pm 0.00014$ \\
$r_{\rm d}h$ & $100.50 \pm 0.53$ & $100.38 \pm 1.01$ & $98.1 \pm 1.2$ & $98.319^{+1.339}_{-1.361}$ & $100.43 \pm 1.01$ & $98.150^{+1.173}_{-1.168}$ & $98.58 \pm 1.20$ \\
$\sigma_8$ & $0.751 \pm 0.029$ & $0.752 \pm 0.029$ & $0.748 \pm 0.021$ & $0.7487^{+0.0286}_{-0.0299}$ & $0.7515 \pm 0.0295$ & $0.7480^{+0.0290}_{-0.0294}$ & $0.749 \pm 0.029$ \\
$w_0$ & \dots & $-0.994 \pm 0.030$ & $ -0.717 \pm 0.089$ & $-0.8197^{+0.1723}_{-0.1519}$ & $-0.9947 \pm 0.030$ & $-0.7703^{+0.0715}_{-0.0720}$ & $ -0.812 \pm 0.075$\\
$w_a$ & \dots & \dots & $-0.97\pm 0.30$ & $0.0348^{+1.1832}_{-1.3867}$ & $0.000003 \pm 0.00002$ & $-0.5026^{+0.1854}_{-0.1640}$ & $-0.385 \pm 0.157$\\
$w_b$ & \dots & \dots & \dots & $-1.7488^{+2.3040}_{-1.9776}$ & \dots & $-0.4259^{+0.1488}_{-0.1411}$ & $-0.374 \pm 0.151$\\
$\alpha$ & \dots & \dots & \dots  & \dots  & $0.018^{+0.101}_{-0.100}$ & $>0$ & $ -0.398 \pm 0.166$  \\
$z_{\rm t}$ & \dots & \dots  & \dots & \dots & unconstrained & $<30$  & $<140$ \\
$z_{\rm eq}$ & \dots & \dots  & \dots & \dots  & \dots & $0.341^{+0.015}_{-0.017}$ & \dots\\
$\gamma$ & \dots & \dots & \dots & \dots & \dots  & $-0.183^{+0.278}_{-0.401}$ & \dots \\
$\beta$ & \dots & \dots & \dots & \dots & \dots & $<0$ &  $ -0.812 \pm 0.075$\\
\hline
$\chi^2_{\rm min}$        & $71.998$ & $71.577$ & $60.942$ & $60.652$ & $71.305$ & $60.558$ & $60.366$\\
$\chi^2_{\rm red}$        & $0.8889$ & $0.8947$ & $0.7714$ & $0.7776$ & $0.9026$ & $0.7716$ & $0.7699$\\
$\Delta\chi^2_{\rm min}$  & $0$ & $-0.421$ & $-11.056$ & $-11.346$ & $-0.693$ & $-11.440$ & $-11.632$ \\
AIC                       & $81.998$ & $83.577$ & $74.942$ & $76.652$ & $85.305$ & $76.558$ & $76.366$ \\
$\Delta$AIC               & $0$ & $1.579$ & $-7.056$ & $-5.346$ & $3.307$ & $-5.440$ & $-5.632$\\
BIC                       & $94.270$ & $98.303$ & $92.123$ & $96.287$ & $102.486$ & $96.193$ & $96.001$\\
$\Delta$BIC               & $0$ & $4.033$ & $-2.147$ & $2.017$ & $8.216$ & $1.923$ & $1.731$\\
\hline
\multicolumn{8}{c}{\textbf{+DESY5}}\\
\hline
$\Omega_{m0}$ & $0.2784 \pm 0.0021$ & $0.2783 \pm 0.0022$ & $0.2989 \pm 0.0043$ & $0.2973^{+0.0047}_{-0.0047}$ & $0.2793 \pm 0.0023$ & $0.2982 \pm 0.0042$ & $0.2975 \pm 0.0041$  \\
$h$ & $0.7046 \pm 0.0016$ & $0.7066 \pm  0.0027$ & $0.6885 \pm 0.0041$ & $0.6908^{+0.0046}_{-0.0046}$  & $0.7070 \pm 0.0026$ & $0.6896 \pm 0.0040$ & $0.6901 \pm 0.0038$ \\
$\omega_{b}$ & $0.02298 \pm 0.00010$ & $0.02290 \pm 0.00013$ & $0.02261 \pm 0.00014$ & $0.02260^{+0.00014}_{-0.00014}$ & $0.02285 \pm 0.00015$ & $0.02261 \pm 0.00014$ & $0.02263 \pm 0.00013$ \\
$r_{\rm d}h$ & $102.98 \pm 0.27$ & $103.38 \pm 0.52$ & $100.74 \pm 0.68$ & $100.99^{+0.71}_{-0.71}$ & $103.48 \pm 0.51$ & $100.79 \pm 0.67$ & $100.83 \pm 0.64$ \\
$\sigma_8$ & $0.783 \pm 0.030$ & $0.781 \pm 0.030$ & $0.77 \pm 0.021$ & $0.769^{+0.030}_{-0.029}$ & $0.778 \pm 0.030$ & $0.769 \pm 0.030$ & $0.774 \pm 0.030$ \\
$w_0$ & $\cdots$ & $-1.019 \pm 0.021$ & $ -0.721 \pm 0.057$ & $-0.833^{+0.117}_{-0.108}$ & $-1.022 \pm 0.018$ & $-0.783\pm 0.048 $ & $ -0.794 \pm 0.047$\\
$w_a$ & $\cdots$ & $\cdots$ & $-1.17^{+0.24}_{-0.22}$ & $-0.018^{+0.953}_{-1.067}$ & $-0.000005^{+0.000004}_{-0.009434}$ & $-0.603 \pm 0.150$ & $-0.543 \pm 0.130$\\
$w_b$ & $\cdots$ & $\cdots$ & $\cdots$ & $-2.018^{+1.840}_{-1.709}$ & $\cdots$ & $-0.519 \pm 0.11$ & $-0.519 \pm 0.115$\\
$\alpha$ & $\cdots$ & $\cdots$ & $\cdots$  & $\cdots$  & $-0.244^{+0.28}_{-0.091}$ & $>2$ & $ -0.570 \pm 0.156$  \\
$z_{\rm t}$ & $\cdots$ & $\cdots$  & $\cdots$ & $\cdots$ & unconstrained & $<25$  & $<65$ \\
$z_{\rm eq}$ & $\cdots$ & $\cdots$  & $\cdots$ & $\cdots$  & $\cdots$ & $0.379\pm{0.011}$ & $\cdots$\\
$\gamma$ & $\cdots$ & $\cdots$ & $\cdots$ & $\cdots$ & $\cdots$  &$ -0.1^{+0.21}_{-0.33}$ & $\cdots$ \\
$\beta$ & $\cdots$ & $\cdots$ & $\cdots$ & $\cdots$ & $\cdots$ & $<-2$ &  $-0.794 \pm 0.047$\\
\hline
$\chi^2_{\rm min}$        & $1731.10$ & $1728.36$ & $1695.00$ & $1694.28$ & $1720.65$ & $1693.55$ & $1693.21$\\
$\chi^2_{\rm red}$        & $0.916$ & $ 0.916$ & $ 0.898$ & $ 0.898$ & $0.911$ & $ 0.898$ & $0.898$\\
$\Delta\chi^2_{\rm min}$            & $0$ & $-2.740$ & $-36.100$ & $-36.820$ & $-10.450$ & $-37.550$ & $-37.890$\\
AIC                       & $1741.10$ & $1740.36$ & $1709.00$ & $1710.28$ & $1734.65$ & $1709.55$ & $1709.21$ \\
$\Delta$AIC               & $0$ & $-0.74$ & $-32.10$ & $-30.82$ & $-6.45$ & $-31.55$ & $-31.89$\\
BIC                       & $1768.83$ & $1773.63$ & $1747.72$ & $1754.65$ & $1772.53$ & $1753.91$ & $1753.58$\\
$\Delta$BIC               & $0$ & $+4.80$ & $-21.11$ & $-14.18$ & $+3.70$ & $-14.92$ & $-15.25$\\
\hline\hline
\end{tabular}}
\end{center}
\end{table*}

For the $\Lambda$CDM scenario, we vary six cosmological parameters, 
$\{\Omega_{\rm m0},\, h,\, \omega_{\rm b},\, r_{\rm d}h,\, \sigma_8,\, M\}$, 
where $\omega_{\rm b}=\Omega_{\rm b0}h^2$ is the physical baryon density, 
$r_{\rm d}$ denotes the comoving sound horizon at the baryon drag epoch, 
and $M$ is the absolute magnitude of SNeIa. We impose uniform priors on these parameters within the ranges
\begin{align}
&\{\Omega_{\rm m0},\, h,\, \omega_{\rm b},\, r_{\rm d}h,\, \sigma_8,\, M\} 
\in \{[0.2,0.5],\, [0.5,0.8], \nn \\ & [0.005,0.05],\, [60,140],\, [0.5,1.0],\, [-22,-15]\}. \nn
\end{align}
For the other models, these cosmological parameters are kept fixed to the same uniform priors.

The $w$CDM model introduces one additional parameter, the constant dark energy EoS parameter $w_0$, for which we consider a uniform prior over the interval $[-2,1]$. The CPL model further extends the $w$CDM scenario by allowing for time evolution in the EoS through two parameters, $w_0$ and $w_a$, with uniform priors given by $\{w_0,w_{\rm a}\} \in \{[-2,0],\; [-2,1]\}$. CPL-$w_{\rm b}$ has one more parameter $w_{\rm b}$ and we use the uniform prior $\{w_0,w_{\rm a},w_{\rm b}\} \in \{[-2,0],\; [-5,3],\; [-5,5]\}$.

Similar to CPL, mAH model also includes two free parameters, $\alpha$ and $\Omega_\delta$. $\alpha>0$ represents quintessence behaviour while $\al<0$ represents phantom behaviour. We consider uniform priors $\{\alpha,\; \Omega_\delta \}\in \{[-1, 1],\; [0, 1000]\}$. 

The MmAH1 model introduces three additional free parameters compared to $\Lambda$CDM, $\beta$, $\gamma$, and $\Omega_\delta$ (or equivalently $z_{\rm t}$), with $\gamma=\alpha+\beta$. The parameter $\Omega_\delta$ is related to the matter-dark energy equality redshift, $z_{\rm eq}$, through the relation~\eqref{eq:Omega_delta_zeq}. From an observational perspective, $z_{\rm eq}$ is typically much smaller than unity and can be tightly constrained. In contrast, $z_{\rm t}$, is more difficult to determine, as current data are not sufficiently sensitive to this epoch. Motivated by this, we adopt $z_{\rm eq}$ as a model parameter in place of $z_{\rm t}$. Uniform priors on the free parameters are imposed as $\{\beta,\, \gamma,\, z_{\rm eq}\} \in \{[-15,1],\, [-2,1],\, [-0.9,5]\}.$

The free parameters of the MmAH2 model are $\alpha$, $\beta$, and $z_{\rm t}$. 
In this case, we do not consider $z_{\rm eq}$ as a free parameter in place of $z_{\rm t}$, since there is no analytic relation between them as in the MmAH1 case. 
Nevertheless, $z_{\rm eq}$ can always be computed numerically. 
We impose uniform priors on the free parameters as $\{\alpha,\, \beta,\, z_{\rm t}\} \in \{[-2,1],\, [-2,1],\, [0,500]\}$.

In our analysis, parameter constraints are obtained using a Markov Chain Monte Carlo (MCMC) framework implemented with the publicly available {\tt EMCEE} sampler \citep{Foreman-Mackey:2012any}. The resulting posterior chains are analysed and visualised using the {\tt GetDist} package \citep{Lewis:2019xzd}. Convergence of the MCMC chains is assessed using the Gelman–Rubin statistic \citep{Gelman:1992zz}, with the requirement $|R-1|\lesssim0.01$ to ensure reliable sampling of the posterior distribution. Since directly extracting best-fit parameters from MCMC chains can be problematic \cite{Hogg:2017akh}, we also determine the best-fit parameters by explicitly minimizing the total $\chi^2$ using a dedicated optimizer ({\tt scipy.optimize.minimize}) \cite{Virtanen2020SciPy}.

\section{Observational Data}
\label{sec:data}
Our analysis combines multiple cosmological probes, including CMB, BAO, SNeIa, Hubble parameter measurements, and RSD. For the CMB, we use distance priors reconstructed from the Planck 2018 TT, TE, EE+lowE data \cite{Planck:2018vyg,2019JCAP}, along with the Planck measurement of the baryon density $\omega_{b}$ \cite{Planck:2018vyg}. The BAO data are taken from DESI DR2 \cite{DESI:2025zgx}, which include over 14 million extragalactic objects—emission line galaxies (ELGs), luminous red galaxies (LRGs), quasars (QSOs) \cite{DESI:2025qqy}, and Ly$\alpha$ forest tracers \cite{DESI:2025zpo}. These data provide measurements of $D_{M}/r_{d}$ and $D_{H}/r_{d}$ over $0.4 < z < 4.2$, and a low-redshift $D_{V}/r_{d}$ measurement for $0.1 < z < 0.4$, encompassing both isotropic and anisotropic BAO analyses. 
For SNeIa, we adopt the Pantheon$+$ compilation containing 1550 luminosity distance measurements in the range $0.001 < z < 2.26$ \cite{Scolnic:2021amr,Brout:2022vxf}. We also use the 22 binned data points from the Union3 compilation \cite{Rubin:2023ovl}, which includes 2087 cosmologically useful SNeIa, together with the DESY5 sample \cite{DES:2024jxu}, comprising 1635 SNeIa in the redshift range $0.10 < z < 1.13$, along with an additional 194 low-redshift SNeIa ($0.025 < z < 0.10$), resulting in a total of 1829 data points.

\begin{figure*}[ht]
\centering
\includegraphics[scale=0.4]{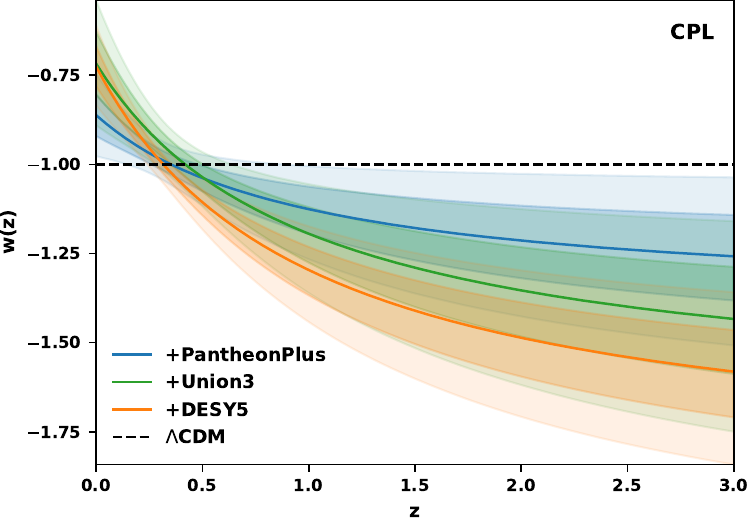}~~~~~~~~~~~~
\includegraphics[scale=0.4]{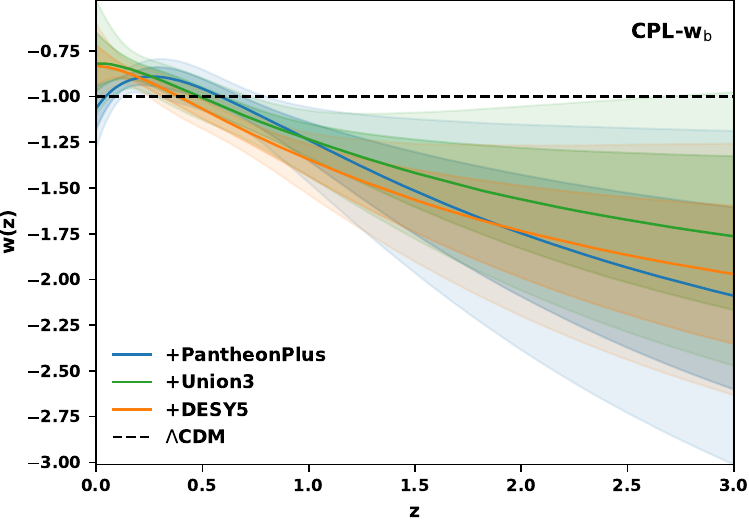} \\
\includegraphics[scale=0.4]{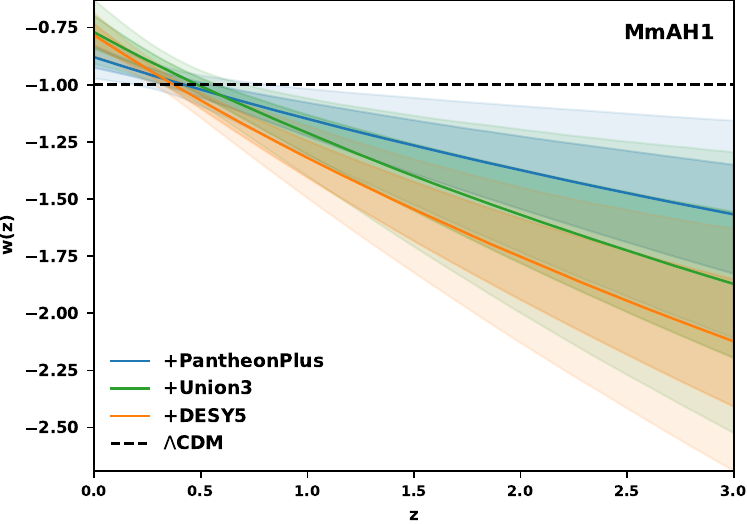}~~~~~~~~~~~
\includegraphics[scale=0.4]{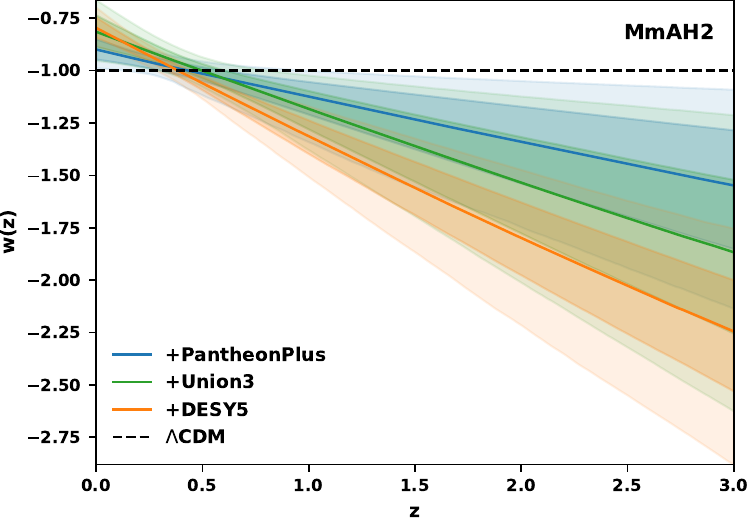}
\caption{Evolution of the EoS  for the CPL (top left), CPL-$w_{\rm b}$ (top right), MmAH1 (bottom left) and MmAH2 (bottom right) models. The solid blue curves
correspond to the median value of the EoS.
The dark colour and light colour shaded regions show the corresponding $1\sigma$ and $2\sigma$ errors.}
\label{fig:wz_error}
\end{figure*}

\begin{figure*}[ht]
\centering
\includegraphics[scale=0.5]{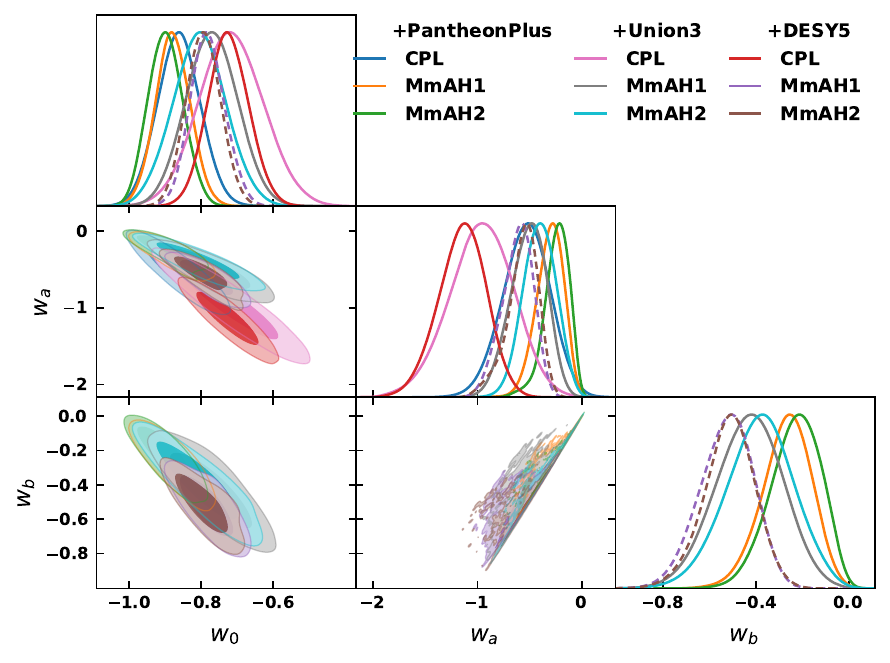}~~~~~~~~~~~~~
\includegraphics[scale=0.5]{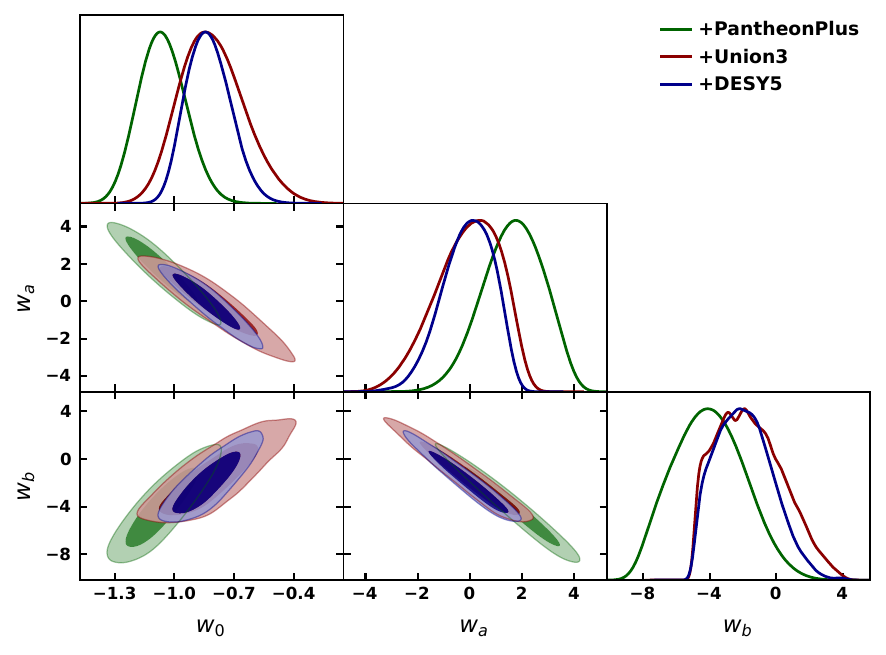}
\caption{$1\sigma$ and $2\sigma$ confidence contours for the parameters $w_0$, $w_a$, and $w_b$ in the CPL, MmAH1, and MmAH2 models (left), and in the CPL-$w_{\rm b}$ model (right).
}
\label{fig:w0-wa-wb}
\end{figure*}

\begin{figure*}[ht]
\centering
\includegraphics[scale=0.5]{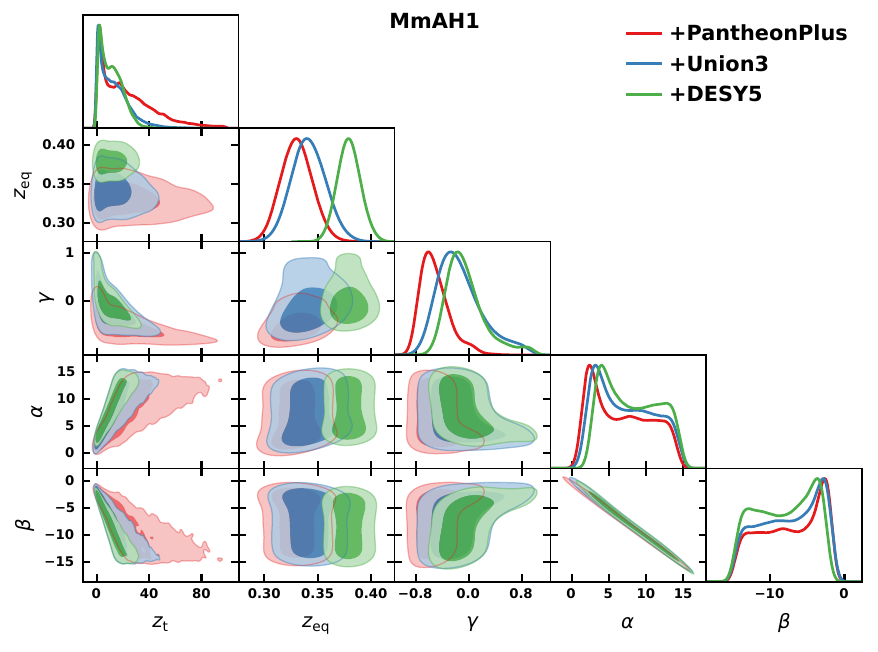}~~~~~~~~~~~~~~~~~
\includegraphics[scale=0.28]{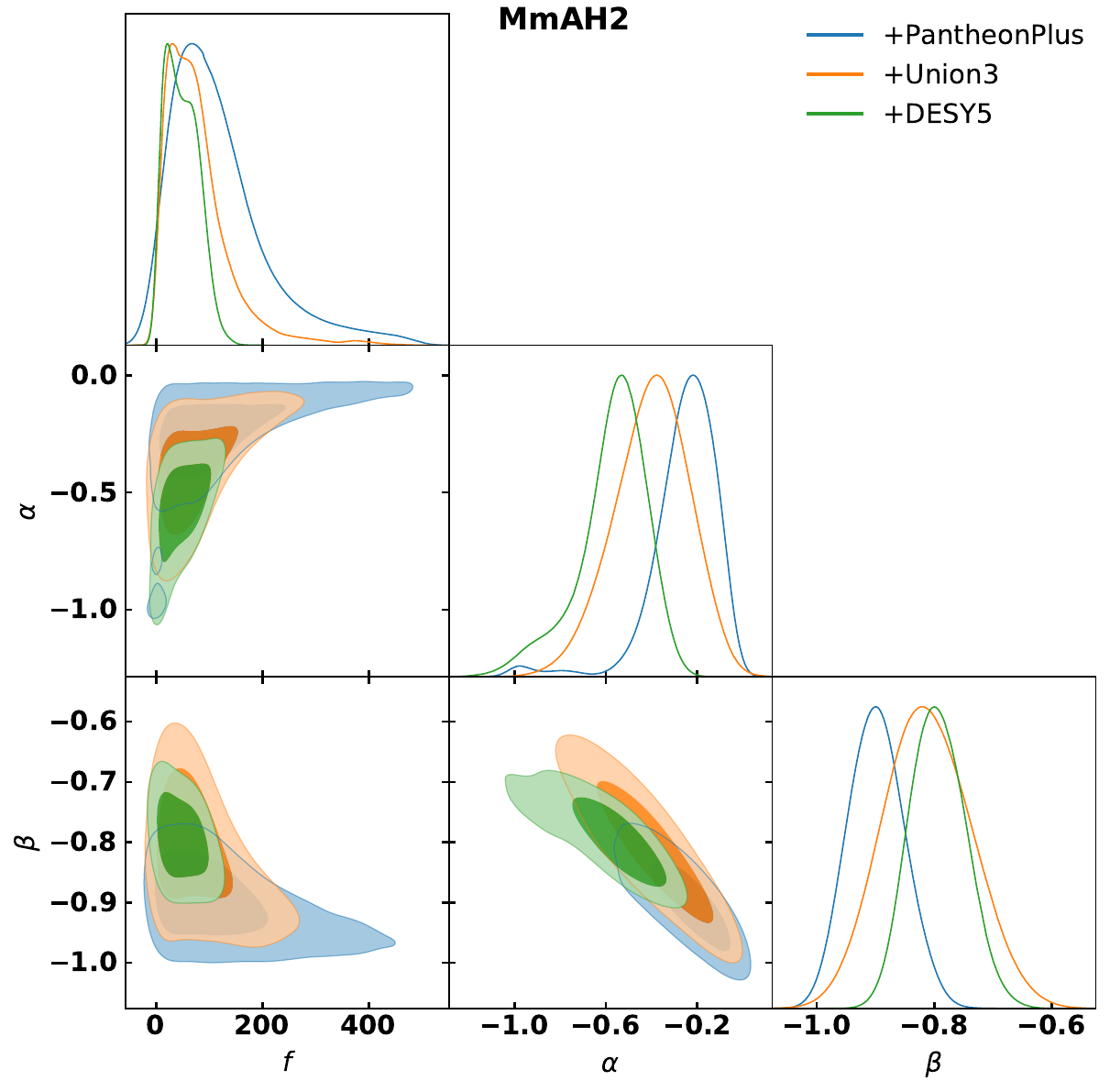}~~~~~~~~~~~~~~~~~
\caption{$1\sigma$ and $2\sigma$ confidence contours for the parameters $z_{\rm t}$, $\alpha$, and $\beta$ in the MmAH1 (left) and MmAH2 (right) models. The contours for $z_{\rm eq}$ and $\gamma$ are also shown for the MmAH1 model.
}
\label{fig6}
\end{figure*}

We further include 31 $H(z)$ measurements spanning $0.07 \leq z \leq 1.965$ from the cosmic chronometer method \cite{2018JCAP...04..051G}. Finally, we include redshift space distortion (RSD) measurements of the growth rate $f\sigma_{8}(z)$ compiled in Ref.~\cite{Nesseris:2017vor}. The growth rate is defined as $f(a)=d\ln\delta/d\ln a$, while the redshift evolution of $\sigma_8$ is given by $\sigma_8(z)=\sigma_8\,\delta(z)/\delta(0)$. Following Ref.~\cite{Nesseris:2017vor}, we construct the covariance matrix and define the $\chi^2$ statistic for the RSD data. Since RSD measurements depend on the fiducial cosmology used by each survey to convert redshifts into distances, we correct for this dependence through the standard Alcock--Paczy\'nski factor
\begin{equation}
\mathrm{ratio}(z)=\frac{H(z)D_A(z)}{H_{\rm fid}(z)D_{A,\rm fid}(z)},
\end{equation}
where the fiducial cosmology corresponds to a flat $\Lambda$CDM model as adopted by the original surveys. The RSD likelihood is then constructed as $\chi^2_{\rm RSD}=V^i C^{-1}_{ij} V^j$, with
$V^i=f\sigma_{8,i}-\mathrm{ratio}(z_i)\,f\sigma_8(z_i,\boldsymbol{\theta})$, where $f\sigma_{8,i}$ is the value of the $i$th data point at the redshift $z_i$ and $f\sigma_8(z_i,\boldsymbol{\theta})$ is the theoretical prediction. Correlations are included for the WiggleZ measurements, while the remaining data points are treated as uncorrelated.

\vspace{0.5cm}
We consider three data combinations corresponding to different SNeIa samples. The combination CMB$+$DESI~DR2$+$H(z)$+$RSD$+$PantheonPlus is denoted as $+$PantheonPlus. Similarly, CMB$+$DESI~DR2$+$H(z)$+$RSD$+$Union3 is referred to as $+$Union3, and CMB $+$ DESI~DR2$+$H(z)$+$RSD$+$DESY5 as $+$DESY5.

\section{Observational Constraints}
\label{sec:results}
Table~\ref{tab:constraints_combined} summarizes the observational constraints on the cosmological parameters for the seven dark energy models considered in this work: $\Lambda$CDM, $w$CDM, CPL, CPL-$w_{\rm b}$, mAH, MmAH1, and MmAH2. For each model, we report the best-fit values with $1\sigma$ uncertainties, along with the minimum chi-squared ($\chi^2_{\rm min}$), AIC, and BIC, together with their differences relative to $\Lambda$CDM. The evolution of the dark energy EoS $w(z)$ with $1\sigma$ and $2\sigma$ confidence regions is shown in Fig.~\ref{fig:wz_error} for CPL (top left), CPL-$w_{\rm b}$ (top right), MmAH1 (bottom left), and MmAH2 (bottom right).

The six standard cosmological parameters are found to be robust across all models for each data combination, with variations well within $1\sigma$. Extending $\Lambda$CDM to $w$CDM introduces one additional parameter, $w_0$. For all three datasets, $w_0$ remains close to $-1$ ($w_0 = -0.991\pm 0.025$, $-0.994\pm 0.030$, and $-1.019\pm 0.021$ for $+$PantheonPlus, $+$Union3 and $+$DESY5, respectively), indicating no significant deviation from the CC. Consequently, the $w$CDM model remains statistically comparable to, but not favoured over, $\Lambda$CDM.

The CPL model allows for time variation in $w(z)$ through parameters $(w_0, w_{\rm a})$. For $+$PantheonPlus, we find $w_0=-0.862\pm0.059$ and $w_{\rm a}=-0.53^{+0.24}_{-0.21}$, consistent with dynamical behaviour of dark energy crossing the phantom divide. The improvement in fit ($\Delta\chi^2=-5.98$) corresponds to $\Delta{\rm AIC}=-1.98$, though $\Delta{\rm BIC}=+8.84$ still penalizes the additional complexity. For $+$Union3, the deviation is stronger with $w_0=-0.717\pm0.089$ and $w{\rm a}=-0.97\pm0.30$, giving $\Delta{\rm AIC}=-7.1$ and $\Delta{\rm BIC}\approx-2.1$, indicating mild statistical preference for CPL. For $+$DESY5, CPL significantly improves the fit ($\Delta\chi^2=-36.1$, $\Delta{\rm AIC}=-32.1$) with $w_0=-0.721\pm0.057$ and $w_{\rm a}=-1.17^{+0.24}_{-0.22}$, showing strong evidence for evolving dark energy.

Adding an extra parameter $w_{\rm b}$ (CPL-$w_{\rm b}$) increases the flexibility of $w(z)$ at low redshifts. For $+$PantheonPlus, $w_0=-1.06^{+0.11}_{-0.13}$, $w{\rm a}=1.7^{+1.3}_{-1.1}$, and $w{\rm b}=-4.0\pm2.2$ capture non-trivial evolution, but statistical tests ($\Delta{\rm AIC}=2.79$, $\Delta{\rm BIC}=19.0$) disfavour the model.
In contrast, for $+$Union3 and $+$DESY5, CPL-$w_{\rm b}$ achieves substantial improvements ($\Delta\chi^2\approx-11.3$ and $-36.8$, respectively) and remains competitive under AIC, though the BIC continues to penalize its complexity.

The mAH model, characterized by $(\alpha, z_{\rm t})$, remains weakly constrained in all datasets. For $+$PantheonPlus, $\alpha=0.027^{+0.076}_{-0.089}$ and $z_{\rm t}$ is unconstrained, reverting effectively to a constant-$w$ behaviour. The small improvement in $\chi^2$ ($\Delta\chi^2=-0.58$) combined with $\Delta{\rm AIC}=3.4$ and $\Delta{\rm BIC}=14.2$ shows that mAH is statistically disfavoured. Similar conclusions hold for $+$Union3 and $+$DESY5, with $\alpha$ consistent with zero and large uncertainty on $z_{\rm t}$, suggesting that the model does not yield meaningful departures from $\Lambda$CDM.

Both MmAH extensions yield significantly improved fits relative to $\Lambda$CDM across all datasets, particularly for DESY5.
For $+$PantheonPlus, MmAH1 gives $w_0=-0.878\pm0.048$, $w_a=-0.304\pm0.124$, and $w_b=-0.272\pm0.106$, with $\Delta\chi^2=-7.16$, $\Delta{\rm AIC}=-1.16$, and $\Delta{\rm BIC}=15.06$. The corresponding auxiliary parameters are $z_{\rm eq}=0.330\pm 0.015$, $\gamma=-0.52^{+0.16}_{-0.26}$, $\alpha>0$, $\beta<0$, and $z{\rm t}<40$.
The MmAH2 model yields comparable improvements ($\Delta\chi^2=-7.12$, $\Delta{\rm AIC}=-1.12$) with $w_0=-0.899\pm0.048$, $w_a=-0.241^{+0.14}_{-0.092}$, and $w_b=-0.230^{+0.12}_{-0.093}$, together with $\alpha=-0.257^{+0.15}_{-0.083}$ and $z{\rm t}<134$. For $+$Union3, both MmAH models again outperform $\Lambda$CDM with $\Delta\chi^2\approx-11.5$ and $\Delta{\rm AIC}\approx-5.5$, while maintaining similar BIC penalties. For $+$DESY5, the improvement is most pronounced, with $\Delta\chi^2\approx-37.6$ and $\Delta{\rm AIC}\approx-31.7$ for both MmAH1 and MmAH2, marking these models almost as good as the CPL parametrization in terms of $\Delta\chi^2$ and $\Delta$AIC.

\section{Evidence of DDE and Quantifying Tension}
\label{sec:tension}

To quantify the statistical disagreement (tension) between two models, A and B, we employ the Mahalanobis distance squared statistic \cite{Mahalanobis:1936,McLachlan:1999,Stephenson1997,Iwamura2002}. 
Let $\boldsymbol{\theta} = \{\theta_1, \dots, \theta_k\}$ denote the vector of $k$ common cosmological parameters, with mean vectors $\boldsymbol{\mu}_i$ and covariance matrices $\mathbf{C}_i$ for $i \in \{\mathrm{A},\mathrm{B}\}$, obtained from the respective MCMC posteriors. 
Assuming the parameter estimates from models A and B are independent Gaussian constraints on the same parameters, the covariance of their difference is
\begin{equation}
\mathbf{C} = \mathbf{C}_A + \mathbf{C}_B.
\end{equation}
The N-dimensional tension is then defined as the squared Mahalanobis distance,
\begin{equation}
\Delta\chi^2(\boldsymbol{\theta}) = (\boldsymbol{\mu}_A - \boldsymbol{\mu}_B)^{\text{T}} \mathbf{C}^{-1} (\boldsymbol{\mu}_A - \boldsymbol{\mu}_B).
\end{equation}

We use the Mahalanobis distance as an approximate way to describe how the mean parameter values shift between models. Interpreting this distance as a chi-squared statistic assumes that the posterior distributions are approximately Gaussian, which is not always the case. In such situations, the quoted tension values should be understood as qualitative indicators rather than as strict measures of statistical significance.

\subsection*{Connection to Wilks' theorem}

In the Gaussian limit, the likelihood around its maximum can be approximated as
\begin{equation}
-2\ln L(\boldsymbol{\theta}) \simeq -2\ln L_{\max} + (\boldsymbol{\theta}-\hat{\boldsymbol{\theta}})^{\mathrm{T}}\mathbf{C}^{-1}(\boldsymbol{\theta}-\hat{\boldsymbol{\theta}}),
\end{equation}
where $L_{\max} \equiv L(\hat{\boldsymbol{\theta}})$ is the maximum likelihood, $\hat{\boldsymbol{\theta}}$ is the maximum-likelihood estimator, and $\mathbf{C}$ is the covariance matrix.
Wilks’ theorem \cite{Wilks:1938} states that the likelihood-ratio statistic
\begin{equation}
\Lambda = -2\ln\frac{L(\boldsymbol{\theta}_0)}{L(\hat{\boldsymbol{\theta}})}
\end{equation}
follows a $\chi^2$ distribution with $k$ degrees of freedom, where $k$ is the number of free parameters.
Here, $\boldsymbol{\theta}_0$ denotes the parameter values under the null hypothesis (e.g. $\Lambda$CDM).
Substituting the Gaussian form of the likelihood yields
\begin{equation}
\Lambda \simeq (\boldsymbol{\theta}_0 - \hat{\boldsymbol{\theta}})^{\mathrm{T}}\mathbf{C}^{-1}(\boldsymbol{\theta}_0 - \hat{\boldsymbol{\theta}}),
\end{equation}
which is identical in form to the squared Mahalanobis distance squared $\Delta\chi^2(\boldsymbol{\theta})$.
Therefore, in the Gaussian regime,
\begin{equation}
\Lambda \approx \Delta\chi^2,
\end{equation}
establishing their asymptotic equivalence as measures of statistical deviation between two hypotheses or datasets. 
This equivalence holds strictly in the limit of large sample size, where the likelihood surface becomes approximately quadratic and the posterior distribution tends toward a multivariate Gaussian. 
Under these conditions, the difference between the true parameters $\boldsymbol{\theta}_0$ and the estimated means $\hat{\boldsymbol{\theta}}$ follows a $\chi^2_k$ distribution, with $k$ corresponding to the number of independent parameters. 
For finite samples or mildly non-Gaussian posteriors, $\Delta\chi^2$ and $\Lambda$ remain close but not exactly identical, with their deviations quantifying the degree of non-Gaussianity in the parameter space.

\subsection*{Quantifying tension}

Under the null hypothesis that the two models are statistically consistent, $\Delta\chi^2$ follows a $\chi^2$ distribution with $k$ degrees of freedom \cite{Anderson2003,Mardia1979}. 
The statistical significance is quantified by the $p$-value,
\begin{equation}
p = 1 - F_{\chi^2}(\Delta\chi^2; k),
\end{equation}
where $F_{\chi^2}(\Delta\chi^2; k)$ denotes the cumulative distribution function (CDF) of the $\chi^2$ distribution, defined as
\begin{equation}
F_{\chi^2}(x; k) \equiv \int_0^x \frac{1}{2^{k/2}\Gamma(k/2)}\, t^{k/2-1} e^{-t/2}\, dt.
\end{equation}
The result can be conveniently expressed as an equivalent Gaussian tension,
\begin{equation}
\sigma_{\rm eq} = \Phi^{-1}\!\left(1 - \frac{p}{2}\right),
\end{equation}
where $\Phi^{-1}$ is the inverse CDF of the standard normal distribution.

\subsection*{Numerical stability and degeneracies}

In practice, the covariance $\mathbf{C}$ may be nearly singular or ill-conditioned because of strong parameter correlations, resulting in very small eigenvalues that cause unstable or spuriously large values of $\Delta\chi^2$. To regularize this, we perform an eigenvalue decomposition,
\begin{equation}
\mathbf{C} = \mathbf{V}\mathbf{\Lambda}\mathbf{V}^{\text{T}},
\end{equation}
where $\mathbf{\Lambda} = \text{diag}(\lambda_1,\dots,\lambda_k)$. 
Eigenmodes with $\lambda_i < \epsilon\,\lambda_{\max}$ (for $\epsilon \ll 1$) are discarded, equivalent to using a pseudo-inverse \cite{Iwamura2002,Tegmark:1996bz}. 
The robust Mahalanobis distance then reads
\begin{equation}
\Delta\chi^2 = (\mathbf{V}_{\rm keep}^{\text{T}}\Delta\boldsymbol{\mu})^{\text{T}} \mathbf{\Lambda}_{\rm keep}^{-1} (\mathbf{V}_{\rm keep}^{\text{T}}\Delta\boldsymbol{\mu}), 
\quad {\rm dof} = N_{\rm keep},
\end{equation}
where $N_{\rm keep}$ is the number of retained eigenmodes.

A complementary diagnostic is the correlation matrix,
\begin{equation}
\mathbf{R} = \text{diag}(\mathbf{C}_A)^{-1/2}\, \mathbf{C}_A\, \text{diag}(\mathbf{C}_A)^{-1/2},
\end{equation}
whose small eigenvalues directly signal strong parameter degeneracies.

\subsection*{Low-dimensional cases}

For subspaces of dimension $k=1,2,3$, the general expression reduces to familiar forms.  

In 1D ($\theta_j$),
\begin{equation}
\Delta\sigma = \frac{|\mu_{A,j} - \mu_{B,j}|}{\sqrt{\sigma^2_{A,j} + \sigma^2_{B,j}}}.
\end{equation}
For 2D and 3D subspaces, the corresponding $2\times2$ or $3\times3$ covariance sub-matrices are used:
\begin{align}
\Delta\chi^2(\theta_1, \theta_2) &= (\boldsymbol{\mu}_A - \boldsymbol{\mu}_B)^{\text{T}} (\mathbf{C}_A^{2\times2} + \mathbf{C}_B^{2\times2})^{-1} (\boldsymbol{\mu}_A - \boldsymbol{\mu}_B), \\
\Delta\chi^2(\theta_1, \theta_2, \theta_3) &= (\boldsymbol{\mu}_A - \boldsymbol{\mu}_B)^{\text{T}} (\mathbf{C}_A^{3\times3} + \mathbf{C}_B^{3\times3})^{-1} (\boldsymbol{\mu}_A - \boldsymbol{\mu}_B),
\end{align}
with corresponding $p$-values and $\sigma_{\rm eq}$ computed from $\chi^2$ distributions with $k=2$ or $3$ degrees of freedom.

\begin{table*}[ht]
\centering
\caption{Local statistical tensions between each model and $\Lambda$CDM for the three data combination. For the 1D case we report the per parameter tension for $w_0$ in the format 
($\Delta\chi^2$, $\sqrt{\Delta\chi^2}$, $p$, $\sigma_{\rm eq}$). 
For the 2D and 3D cases we show the Mahalanobis distance $\Delta\chi^2$ with the corresponding 
($\sqrt{\Delta\chi^2}$, $p$, $\sigma_{\rm eq}$), along with the correlation matrices in the relevant parameter subspaces.}
\label{tab:tension_all}
\resizebox{\textwidth}{!}{%
\begin{tabular}{l @{\hskip 0.7cm} c @{\hskip 0.7cm} c @{\hskip 0.7cm} c}
\hline\hline
Model & 1D: $w_0$ & 2D: $(w_0,w_a)$ & 3D: $(w_0,w_a,w_b)$ \\
\hline
\multicolumn{4}{c}{\textbf{$+$PantheonPlus}} \\
\hline
\multirow{3}{*}{CPL} 
 & $\Delta\chi^2 = 5.43$, $\sqrt{\Delta\chi^2}=2.33$ & $\Delta\chi^2 = 5.81$, $\sqrt{\Delta\chi^2}=2.41$ & $\cdots$ \\
 & $p=0.0198$, $\sigma_{\rm eq}=2.33\sigma$ & $p=0.0546$, $\sigma_{\rm eq}=1.92\sigma$ & $\cdots$ \\
 & $\cdots$ & Corr$_{2D}=\begin{pmatrix}1 & -0.885 \\ -0.885 & 1\end{pmatrix}$ & $\cdots$ \\
\hline
\multirow{3}{*}{CPL-$w_b$}
 & $\Delta\chi^2 = 0.27$, $\sqrt{\Delta\chi^2}=0.52$ & $\Delta\chi^2 = 7.64$, $\sqrt{\Delta\chi^2}=2.76$ & $\Delta\chi^2 = 8.29$, $\sqrt{\Delta\chi^2}=2.88$ \\
 & $p=0.603$, $\sigma_{\rm eq}=0.52\sigma$ & $p=0.0220$, $\sigma_{\rm eq}=2.29\sigma$ & $p=0.0404$, $\sigma_{\rm eq}=2.05\sigma$ \\
 & $\cdots$ & Corr$_{2D}=\begin{pmatrix}1 & -0.938 \\ -0.938 & 1\end{pmatrix}$ &
   Corr$_{3D}=\begin{pmatrix}
   1 & -0.938 & 0.859 \\
   -0.938 & 1 & -0.976 \\
   0.859 & -0.976 & 1
   \end{pmatrix}$ \\
\hline
\multirow{3}{*}{MmAH1}
 & $\Delta\chi^2 = 6.34$, $\sqrt{\Delta\chi^2}=2.52$ & $\Delta\chi^2 = 6.58$, $\sqrt{\Delta\chi^2}=2.56$ & $\Delta\chi^2 = 7.03$, $\sqrt{\Delta\chi^2}=2.65$ \\
 & $p=0.0118$, $\sigma_{\rm eq}=2.52\sigma$ & $p=0.037$, $\sigma_{\rm eq}=2.08\sigma$ & $p=0.071$, $\sigma_{\rm eq}=1.81\sigma$ \\
 & $\cdots$ & Corr$_{2D}=\begin{pmatrix}1 & -0.803 \\ -0.803 & 1\end{pmatrix}$ &
   Corr$_{3D}=\begin{pmatrix}
   1 & -0.803 & -0.807 \\
   -0.803 & 1 & 0.919 \\
   -0.807 & 0.919 & 1
   \end{pmatrix}$ \\
\hline
\multirow{3}{*}{MmAH2}
 & $\Delta\chi^2 = 4.51$, $\sqrt{\Delta\chi^2}=2.12$ & $\Delta\chi^2 = 4.71$, $\sqrt{\Delta\chi^2}=2.17$ & $\Delta\chi^2 = 5.03$, $\sqrt{\Delta\chi^2}=2.24$ \\
 & $p=0.0337$, $\sigma_{\rm eq}=2.12\sigma$ & $p=0.095$, $\sigma_{\rm eq}=1.67\sigma$ & $p=0.17$, $\sigma_{\rm eq}=1.37\sigma$ \\
 & $\cdots$ & Corr$_{2D}=\begin{pmatrix}1 & -0.799 \\ -0.799 & 1\end{pmatrix}$ &
   Corr$_{3D}=\begin{pmatrix}
   1 & -0.799 & -0.818 \\
   -0.799 & 1 & 0.961 \\
   -0.818 & 0.961 & 1
   \end{pmatrix}$ \\
\hline
\multicolumn{4}{c}{\textbf{$+$Union3}} \\
\hline
\multirow{3}{*}{CPL} 
 & $\Delta\chi^2 = 10.01$, $\sqrt{\Delta\chi^2}=3.16$ & $\Delta\chi^2 = 10.34$, $\sqrt{\Delta\chi^2}=3.22$ & $\cdots$ \\
 & $p=1.56\times10^{-3}$, $\sigma_{\rm eq}=3.16\sigma$ & $p=5.70\times10^{-3}$, $\sigma_{\rm eq}=2.76\sigma$ & $\cdots$ \\
 & $\cdots$ & Corr$_{2D}=\begin{pmatrix}1 & -0.925 \\ -0.925 & 1\end{pmatrix}$ & $\cdots$ \\
\hline
\multirow{3}{*}{CPL-$w_b$}
 & $\Delta\chi^2 = 1.39$, $\sqrt{\Delta\chi^2}=1.18$ & $\Delta\chi^2 = 9.29$, $\sqrt{\Delta\chi^2}=3.05$ & $\Delta\chi^2 = 10.78$, $\sqrt{\Delta\chi^2}=3.29$ \\
 & $p=0.239$, $\sigma_{\rm eq}=1.18\sigma$ & $p=9.61\times10^{-3}$, $\sigma_{\rm eq}=2.59\sigma$ & $p=1.30\times10^{-2}$, $\sigma_{\rm eq}=2.48\sigma$ \\
 & $\cdots$ & Corr$_{2D}=\begin{pmatrix}1 & -0.929 \\ -0.929 & 1\end{pmatrix}$ &
   Corr$_{3D}=\begin{pmatrix}
   1 & -0.929 & 0.816 \\
   -0.929 & 1 & -0.964 \\
   0.816 & -0.964 & 1
   \end{pmatrix}$ \\
\hline
\multirow{3}{*}{MmAH1}
 & $\Delta\chi^2 = 10.32$, $\sqrt{\Delta\chi^2}=3.21$ & $\Delta\chi^2 = 10.57$, $\sqrt{\Delta\chi^2}=3.25$ & $\Delta\chi^2 = 10.68$, $\sqrt{\Delta\chi^2}=3.27$ \\
 & $p=1.31\times10^{-3}$, $\sigma_{\rm eq}=3.21\sigma$ & $p=5.06\times10^{-3}$, $\sigma_{\rm eq}=2.80\sigma$ & $p=1.36\times10^{-2}$, $\sigma_{\rm eq}=2.47\sigma$ \\
 & $\cdots$ & Corr$_{2D}=\begin{pmatrix}1 & -0.839 \\ -0.839 & 1\end{pmatrix}$ &
   Corr$_{3D}=\begin{pmatrix}
   1 & -0.839 & -0.836 \\
   -0.839 & 1 & 0.836 \\
   -0.836 & 0.836 & 1
   \end{pmatrix}$ \\
\hline
\multirow{3}{*}{MmAH2}
 & $\Delta\chi^2 = 6.35$, $\sqrt{\Delta\chi^2}=2.52$ & $\Delta\chi^2 = 6.58$, $\sqrt{\Delta\chi^2}=2.56$ & $\Delta\chi^2 = 6.61$, $\sqrt{\Delta\chi^2}=2.57$ \\
 & $p=0.0118$, $\sigma_{\rm eq}=2.52\sigma$ & $p=0.0373$, $\sigma_{\rm eq}=2.08\sigma$ & $p=0.0855$, $\sigma_{\rm eq}=1.72\sigma$ \\
 & $\cdots$ & Corr$_{2D}=\begin{pmatrix}1 & -0.887 \\ -0.887 & 1\end{pmatrix}$ &
   Corr$_{3D}=\begin{pmatrix}
   1 & -0.887 & -0.895 \\
   -0.887 & 1 & 0.995 \\
   -0.895 & 0.995 & 1
   \end{pmatrix}$ \\

\hline
\multicolumn{4}{c}{\textbf{$+$DESY5}} \\
\hline
\multirow{3}{*}{CPL} 
 & $\Delta\chi^2 = 23.34$, $\sqrt{\Delta\chi^2}=4.83$ & $\Delta\chi^2 = 25.71$, $\sqrt{\Delta\chi^2}=5.07$ & $\cdots$ \\
 & $p=1.36\times10^{-6}$, $\sigma_{\rm eq}=4.83\sigma$ & $p=2.61\times10^{-6}$, $\sigma_{\rm eq}=4.70\sigma$ & $\cdots$ \\
 & $\cdots$ & Corr$_{2D}=\begin{pmatrix}1 & -0.893 \\ -0.893 & 1\end{pmatrix}$ & $\cdots$ \\
\hline
\multirow{3}{*}{CPL-$w_b$}
 & $\Delta\chi^2 = 2.42$, $\sqrt{\Delta\chi^2}=1.56$ & $\Delta\chi^2 = 18.32$, $\sqrt{\Delta\chi^2}=4.28$ & $\Delta\chi^2 = 23.92$, $\sqrt{\Delta\chi^2}=4.89$ \\
 & $p=0.120$, $\sigma_{\rm eq}=1.56\sigma$ & $p=1.05\times10^{-4}$, $\sigma_{\rm eq}=3.88\sigma$ & $p=2.60\times10^{-5}$, $\sigma_{\rm eq}=4.21\sigma$ \\
 
\end{tabular}}
\end{table*}

\begin{table*}[t]
\centering
\caption*{}
\resizebox{\textwidth}{!}{%
\begin{tabular}{l @{\hskip 0.7cm} c @{\hskip 0.7cm} c @{\hskip 0.7cm} c}
& $\cdots$ & Corr$_{2D}=\begin{pmatrix}1 & -0.938 \\ -0.938 & 1\end{pmatrix}$ &
   Corr$_{3D}=\begin{pmatrix}
   1 & -0.938 & 0.842 \\
   -0.938 & 1 & -0.967 \\
   0.842 & -0.967 & 1
   \end{pmatrix}$ \\
\hline
\multirow{3}{*}{MmAH1}
 & $\Delta\chi^2 = 20.60$, $\sqrt{\Delta\chi^2}=4.54$ & $\Delta\chi^2 = 21.03$, $\sqrt{\Delta\chi^2}=4.58$ & $\Delta\chi^2 = 23.52$, $\sqrt{\Delta\chi^2}=4.85$ \\
 & $p=5.67\times10^{-6}$, $\sigma_{\rm eq}=4.54\sigma$ & $p=2.71\times10^{-5}$, $\sigma_{\rm eq}=4.20\sigma$ & $p=3.14\times10^{-5}$, $\sigma_{\rm eq}=4.16\sigma$ \\
 & $\cdots$ & Corr$_{2D}=\begin{pmatrix}1 & -0.818 \\ -0.818 & 1\end{pmatrix}$ &
   Corr$_{3D}=\begin{pmatrix}
   1 & -0.818 & -0.766 \\
   -0.818 & 1 & 0.788 \\
   -0.766 & 0.788 & 1
   \end{pmatrix}$ \\
\hline
\multirow{3}{*}{MmAH2}
 & $\Delta\chi^2 = 18.81$, $\sqrt{\Delta\chi^2}=4.34$ & $\Delta\chi^2 = 19.78$, $\sqrt{\Delta\chi^2}=4.45$ & $\Delta\chi^2 = 22.86$, $\sqrt{\Delta\chi^2}=4.78$ \\
 & $p=1.44\times10^{-5}$, $\sigma_{\rm eq}=4.34\sigma$ & $p=5.07\times10^{-5}$, $\sigma_{\rm eq}=4.05\sigma$ & $p=4.33\times10^{-5}$, $\sigma_{\rm eq}=4.09\sigma$ \\
 & $\cdots$ & Corr$_{2D}=\begin{pmatrix}1 & -0.841 \\ -0.841 & 1\end{pmatrix}$ &
   Corr$_{3D}=\begin{pmatrix}
   1 & -0.841 & -0.826 \\
   -0.841 & 1 & 0.970 \\
   -0.826 & 0.970 & 1
   \end{pmatrix}$ \\
\hline\hline
\end{tabular}}
\end{table*}

\begin{table*}[ht]
\centering
\caption{Global statistical tensions between $\Lambda$CDM and the models MmAH1 and MmAH2 models with  for the three data combinations.}
\label{tab:tension_MmAH_3D}
\resizebox{\textwidth}{!}{%
\begin{tabular}{l @{\hskip 0.9cm} c @{\hskip 0.9cm} c @{\hskip 0.9cm} c}
\hline\hline
Model 
& $+$PantheonPlus 
& $+$Union3 
& $+$DESY5 \\
\hline
MmAH1 $(\beta,\gamma,z_{\rm eq})$
&
$\begin{array}{c}
(\Delta\chi^2,\sqrt{\Delta\chi^2}) = (10.25,\,3.20) \\[2pt]
(p,\sigma_{\rm eq}) = (0.0166,\,2.40\sigma) \\[4pt]
\text{Corr}_{3D} =
\begin{pmatrix}
1 & 0.35 & -0.03 \\
0.35 & 1 & 0.30 \\
-0.03 & 0.30 & 1
\end{pmatrix}
\end{array}$
&
$\begin{array}{c}
(\Delta\chi^2,\sqrt{\Delta\chi^2}) = (14.76,\,3.84) \\[2pt]
(p,\sigma_{\rm eq}) = (2.03\times10^{-3},\,3.09\sigma) \\[4pt]
\text{Corr}_{3D} =
\begin{pmatrix}
1 & 0.41 & -0.07 \\
0.41 & 1 & 0.22 \\
-0.07 & 0.22 & 1
\end{pmatrix}
\end{array}$
&
$\begin{array}{c}
(\Delta\chi^2,\sqrt{\Delta\chi^2}) = (37.35,\,6.11) \\[2pt]
(p,\sigma_{\rm eq}) = (3.87\times10^{-8},\,5.50\sigma) \\[4pt]
\text{Corr}_{3D} =
\begin{pmatrix}
1 & 0.46 & -0.01 \\
0.46 & 1 & 0.10 \\
-0.01 & 0.10 & 1
\end{pmatrix}
\end{array}$ \\
\hline
MmAH2 $(\beta,\alpha,z_t)$
&
$\begin{array}{c}
(\Delta\chi^2,\sqrt{\Delta\chi^2}) = (1.53,\,1.24) \\[2pt]
(p,\sigma_{\rm eq}) = (0.215,\,1.24\sigma) \\[4pt]
\text{Corr}_{3D} =
\begin{pmatrix}
1 & -0.70 & -0.45 \\
-0.70 & 1 & 0.54 \\
-0.45 & 0.54 & 1
\end{pmatrix}
\end{array}$
&
$\begin{array}{c}
(\Delta\chi^2,\sqrt{\Delta\chi^2}) = (1.48,\,1.22) \\[2pt]
(p,\sigma_{\rm eq}) = (0.223,\,1.22\sigma) \\[4pt]
\text{Corr}_{3D} =
\begin{pmatrix}
1 & -0.87 & -0.46 \\
-0.87 & 1 & 0.58 \\
-0.46 & 0.58 & 1
\end{pmatrix}
\end{array}$
&
$\begin{array}{c}
(\Delta\chi^2,\sqrt{\Delta\chi^2}) = (33.78,\,5.81) \\[2pt]
(p,\sigma_{\rm eq}) = (4.63\times10^{-8},\,5.47\sigma) \\[4pt]
\text{Corr}_{3D} =
\begin{pmatrix}
1 & -0.81 & -0.36 \\
-0.81 & 1 & 0.57 \\
-0.36 & 0.57 & 1
\end{pmatrix}
\end{array}$ \\
\hline\hline
\end{tabular}}
\end{table*}

\subsection*{Cosmological Tension}
Table~\ref{tab:tension_all} and Table~\ref{tab:tension_MmAH_3D} summarise the statistical consistency of various dark energy parametrisations with respect to the $\Lambda$CDM baseline across the $+$PantheonPlus, $+$Union3 and $+$DESY5 data combinations. In Table~\ref{tab:tension_all}, we consider the {\it local tension}, \emph{i.e.}, the tension evaluated at low redshifts. Specifically, we compute the 1D, 2D, and 3D tensions corresponding to $w_0$, $(w_0,w_{\rm a})$, and $(w_0,w_{\rm a},w_{\rm b})$, respectively. Here, $w_0$, $w_{\rm a}$, and $w_{\rm b}$ are the CPL coefficients, obtained by expanding the dark energy EoS $w(z)$ up to second order around $z=0$. The motivation for considering this local tension is to facilitate a direct and consistent comparison of the MmAH models not only with $\Lambda$CDM but also with the CPL and CPL-$w_{\rm b}$ parametrizations.

Unlike the CPL and CPL-$w_{\rm b}$ parametrizations, the MmAH1 and MmAH2 models represent more general forms of $w(z)$ and are valid over the entire redshift range. This {\it global} nature of $w(z)$ in the MmAH1 and MmAH2 models allows us to investigate the corresponding {\it global tension} with the $\Lambda$CDM model by directly comparing the fundamental model parameters, namely $(\beta,\gamma,z_{\rm eq})$ for MmAH1 and $(\beta,\alpha,z_t)$ for MmAH2. The resulting global tension analysis is presented in Table~\ref{tab:tension_MmAH_3D}.

For the $+$PantheonPlus set, the CPL and MmAH1 models show moderate deviations from $\Lambda$CDM, with $\Delta\chi^2\simeq5.4$--$6.3$ and equivalent significances of $2.3\sigma$--$2.5\sigma$ in one dimension. 
The extended CPL-$w_{\rm b}$ model, however, remains statistically consistent with $\Lambda$CDM in one dimension ($\sigma_{\rm eq}=0.5\sigma$), but develops a mild tension in higher-dimensional subspaces ($\Delta\chi^2\simeq8.3$, $p\simeq0.04$). 
The correlation matrices reveal strong parameter degeneracies, especially between $w_a$ and $w_b$ in CPL-$w_{\rm b}$, with $|R_{ij}|\simeq0.97$, indicating limited independent constraining power for higher-order terms.  

For the $+$Union3 combination, all dynamical models display stronger departures from $\Lambda$CDM. 
The CPL and MmAH1 parametrisations reach $\Delta\chi^2\simeq10$--$10.6$ ($p\lesssim10^{-3}$), corresponding to $2.8\sigma$--$3.2\sigma$ tensions in one dimension and $2.5\sigma$--$2.8\sigma$ in higher-dimensional subspaces. 
The MmAH2 model yields slightly lower significance ($\sim2.1\sigma$ in 2D). The correlation strengths among parameters remain high ($|R_{ij}|\sim0.9$), but somewhat reduced compared to $+$PantheonPlus, reflecting improved parameter separability with the Union3 calibration.

The $+$DESY5 dataset amplifies these trends substantially. 
All dynamical models exhibit significant deviations from $\Lambda$CDM, with $\Delta\chi^2\simeq19$--$25$ corresponding to $4.0\sigma$--$5.0\sigma$ tensions. 
The CPL model shows the strongest overall inconsistency ($\sigma_{\rm eq}\approx5.1\sigma$ in 2D), closely followed by MmAH1 and MmAH2 with $\sigma_{\rm eq}\approx4.1\sigma$ in three dimensions. 
These results highlight that the inclusion of $+$DESY5 enhances the sensitivity to dynamical effects in the dark energy sector, providing a significant evidence for deviations from a CC.

A complementary picture emerges from the global tension analysis based on the full parameter spaces of the MmAH models (Table~\ref{tab:tension_MmAH_3D}). For the $+$PantheonPlus and $+$Union3 datasets, MmAH1 shows mild-to-moderate global deviations from $\Lambda$CDM at the $\sim2.4\sigma$ and $\sim3.1\sigma$ levels, respectively, while MmAH2 remains largely consistent with $\Lambda$CDM, with tensions below $\sim1.3\sigma$. In contrast, the inclusion of $+$DESY5 leads to an enhancement of global tensions for both models, yielding highly significant discrepancies of $\sim5.5\sigma$ for MmAH1 and $\sim5.5\sigma$ for MmAH2. The associated correlation matrices indicate non-negligible but model-dependent parameter couplings, suggesting that the strong DESY5-driven tensions are not solely driven by degeneracies but reflect genuine global departures from $\Lambda$CDM in the MmAH framework.

\section{Conclusions}
\label{sec:conc}

In this work, we have introduced and studied two new three-parameter dark energy parametrizations, denoted as MmAH1~\eqref{eq:MmAH} and MmAH2~\eqref{eq:wz-main}, which extend the recently proposed mAH form~\eqref{eq:AH}. These models are designed to provide a smooth, bounded, and physically motivated evolution of the dark energy equation of state (EoS), capable of reproducing a wide class of dynamical behaviours while reducing to $\Lambda$CDM in the appropriate limit. To test their performance, we carried out a comprehensive joint analysis using current cosmological datasets, including the CMB compressed likelihood, DESI DR2 BAO, $H(z)$ measurements, and RSD data, in combination with three SNeIa compilations, PantheonPlus, Union3 and DESY5. These yield three independent data combinations, denoted as $+$PantheonPlus, $+$Union3, and $+$DESY5.

We compared the MmAH models with the $\Lambda$CDM model, its one parameter extension $w$CDM, the two parameter CPL and mAH parametrizations, and the three parameter CPL-$w_b$ model. Parameter constraints and model comparison statistics for all cases are summarised in Table~\ref{tab:constraints_combined}. The six background cosmological parameters remain highly stable across all models and datasets, with variations well within $1\sigma$. 

For the $w$CDM model, we obtain $w_0=-0.991\pm0.025$ ($+$PantheonPlus), $-0.994\pm0.030$ ($+$Union3), and $-1.019\pm0.021$ ($+$DESY5$)$, indicating full consistency with the CC. The mAH parametrization yields results nearly indistinguishable from $w$CDM, with negligible improvement in $\chi^2_{\rm min}$.

The CPL parametrization shows moderate evidence for redshift evolution in the dark energy equation of state (EoS) across all three data combinations. For $+$PantheonPlus, we obtain $w_0=-0.862\pm0.059$ and $w_a=-0.53^{+0.24}_{-0.21}$, indicating a mildly dynamical, non phantom present day EoS that evolves toward the phantom regime at intermediate redshifts. For $+$Union3, the evolution is more pronounced, with $w_0=-0.717\pm0.089$ and $w_a=-0.97\pm0.30$, while $+$DESY5 yields $w_0=-0.721\pm0.057$ and $w_a=-1.17^{+0.24}_{-0.22}$, maintaining the same qualitative trend but with tighter constraints. In terms of model performance, CPL improves the fit over $\Lambda$CDM with $\Delta\chi^2=-5.98$, $-11.06$, and $-36.10$ for $+$PantheonPlus, $+$Union3, and $+$DESY5 respectively. The corresponding information criteria are $\Delta{\rm AIC}=-1.98$, $-7.06$, and $-32.10$, showing a mild to strong preference for CPL, particularly when DESY5 is included. However, the Bayesian Information Criterion yields $\Delta{\rm BIC}=+8.84$, $-2.15$, and $-21.11$ for the same datasets, implying that while the additional parameter $w_a$ significantly improves the likelihood, it is statistically justified only for the higher precision $+$DESY5 combination. Overall, CPL remains one of the most versatile two-parameter extensions of $\Lambda$CDM, capturing possible deviations from a constant EoS.

Although the CPL-$w_b$ model introduces an additional curvature term in $w(z)$ and exhibits strong parameter degeneracies in some parameter subspaces, its statistical performance depends sensitively on the dataset. 
For $+$PantheonPlus the model is penalised by the information criteria ($\Delta{\rm AIC}=+2.79$, $\Delta{\rm BIC}=+19.02$), indicating no improvement over simpler models. 
However, for $+$Union3 the CPL-$w_b$ model achieves a substantially better AIC ($\Delta{\rm AIC}=-5.35$) while the BIC shows only a mild penalty ($\Delta{\rm BIC}=+2.02$), and for $+$DESY5 it is strongly preferred by both criteria ($\Delta{\rm AIC}=-30.82$, $\Delta{\rm BIC}=-14.18$). 
Thus CPL-$w_b$ is disfavoured by the $+$PantheonPlus combination but becomes competitive in the $+$Union3 and $+$DESY5 cases.

The proposed MmAH1 and MmAH2 parametrizations provide consistently improved fits over $\Lambda$CDM across all three data combinations. For $+$PantheonPlus, both models yield $\Delta\chi^2=-7.16$ and $-7.12$ relative to $\Lambda$CDM, comparable to or slightly better than CPL ($\Delta\chi^2=-5.98$), with $\Delta{\rm AIC}\approx -1.1$ and $\Delta{\rm BIC}\approx +15$, indicating modest penalties from the additional parameters. The best-fit EoS parameters are $w_0=-0.878\pm0.048$, $w_a=-0.304\pm0.124$, and $w_b=-0.272\pm0.106$ for MmAH1, and $w_0=-0.899\pm0.048$, $w_a=-0.241^{+0.14}_{-0.09}$, and $w_b=-0.230^{+0.12}_{-0.09}$ for MmAH2, corresponding to a smooth non phantom behaviour at present that transitions toward $w<-1$ at intermediate redshifts. For the $+$Union3 dataset, both extensions achieve a stronger improvement, with $\Delta\chi^2=-11.44$ and $-11.63$, closely matching CPL ($\Delta\chi^2=-11.06$) but with smaller AIC penalties ($\Delta{\rm AIC}\simeq -5.5$ for MmAH versus $-7.1$ for CPL) and nearly identical $\Delta{\rm BIC}\sim2$, indicating comparable statistical support. The most significant enhancement appears for the $+$DESY5 combination, where $\Delta\chi^2=-37.55$ and $-37.89$ for MmAH1 and MmAH2, respectively, outperform CPL ($\Delta\chi^2=-36.10$) and achieving nearly identical likelihood improvement with $\Delta{\rm AIC}$ values ($-31.6$ and $-31.9$ versus $-32.1$) and comparable $\Delta{\rm BIC}$ ($-14.9$ and $-15.3$ versus $-21.1$). These results confirm that the MmAH family provides a statistically competitive alternative to CPL.

The degree of statistical consistency between $\Lambda$CDM and its dynamical extensions has been quantified using the Mahalanobis distance in one, two, and three dimensional parameter subspaces, as summarised in Table~\ref{tab:tension_all} for the local tension, valid at low redshifts. 
For the $+$PantheonPlus combination, all models show mild to moderate deviations from $\Lambda$CDM, typically in the range of $1.5$--$2.5\sigma$. The CPL and MmAH1 parametrizations exhibit the largest shifts, with $\sigma_{\rm eq}\simeq2.3\sigma$ and $2.5\sigma$ respectively, while the CPL-$w_b$ and MmAH2 extensions remain consistent within $2\sigma$. The corresponding correlation matrices indicate strong anti-correlation between $w_0$ and $w_a$ (with $R_{w_0w_a}\sim-0.8$ to $-0.9$) and, for higher-order models, an additional positive correlation between $w_a$ and $w_b$ ($R_{w_aw_b}\gtrsim0.9$), implying partial degeneracy among the EoS parameters.

For $+$Union3, the overall level of tension increases slightly, reaching $2$--$3\sigma$ for all dynamical models. CPL yields $\sigma_{\rm eq}=3.2\sigma$ in 1D and $2.8\sigma$ in 2D, while the MmAH1 and CPL-$w_b$ extensions show comparable deviations of $2.4$--$2.6\sigma$ in 3D space. The MmAH2 parametrization remains the most consistent with $\Lambda$CDM, with $\sigma_{\rm eq}\lesssim2\sigma$ in all subspaces. The correlation structure again reveals strong parameter couplings, with $|R_{w_aw_b}|\simeq0.84$--$0.99$, underscoring the limited sensitivity of current data to higher-order EoS evolution.

The $+$DESY5 dataset amplifies these tensions substantially. CPL shows the largest departure from $\Lambda$CDM, reaching $\sim5\sigma$ in both 1D and 2D spaces, followed closely by MmAH1 and MmAH2 with $\sigma_{\rm eq}\simeq4$--$4.5\sigma$. The CPL-$w_b$ model also exhibits a significant deviation, up to $4.2\sigma$ in the full 3D parameter space. Despite the higher dimensionality, the correlation matrices remain highly structured, with $R_{w_0w_a}\approx-0.9$ and $R_{w_aw_b}\approx0.8$--$0.97$, confirming the persistence of strong degeneracies. Overall, these results suggest that while all dynamical parametrizations exhibit notable departures from $\Lambda$CDM when constrained by high-precision data, the MmAH1 and MmAH2 forms provide a statistically consistent yet flexible description of dark energy evolution, maintaining moderate tensions and well-behaved correlations across datasets.

Importantly, the global tension analysis, shown in Table~\ref{tab:tension_MmAH_3D}, reinforces this conclusion. When the full parameter spaces of the MmAH models are compared directly with $\Lambda$CDM, MmAH1 exhibits a pronounced global inconsistency for the $+$DESY5 combination ($\sim5.5\sigma$), whereas MmAH2 shows a comparatively milder but still significant deviation. For the $+$PantheonPlus and $+$Union3 datasets, the global tensions remain moderate, indicating that the MmAH parametrizations can accommodate both low- and high-redshift information in a controlled manner. This highlights the advantage of globally defined models in assessing departures from $\Lambda$CDM beyond the local, low-redshift regime.    

Overall, our analysis demonstrates that while the $\Lambda$CDM model remains statistically consistent with current cosmological observations, several dynamical dark energy parametrizations provide a moderately improved description of the data. These results indicate that a richer and smoothly evolving dark energy behaviour remains compatible with present data at a statistically non-negligible level. Future high-precision cosmological surveys will be crucial in establishing whether the observed deviations reflect genuine departures from a CC or are merely statistical fluctuations within the $\Lambda$CDM framework.

\begin{acknowledgments}
The authors acknowledge the High Performance Computing facility Pegasus at IUCAA, Pune, India, for providing computing resources. SA acknowledges the financial support received from the Council of Scientific and Industrial Research (CSIR), Government of India, through the CSIR NET-SRF fellowship (File No. 09/0466(12904)/2021). MWH acknowledges the hospitality and support of the Department of Theoretical Physics, CERN, Geneva, Switzerland, where part of this work was carried out during a visit.
\end{acknowledgments}

\bibliographystyle{JHEP} 
\bibliography{references}

\end{document}